\documentclass[preprint,texlive=2011,UKenglish,texmf]{atlasdoc}
\pdfoutput=1
\usepackage{atlaspackage}
\usepackage{cite}

\usepackage{atlasphysics}
\usepackage{rotating, graphicx}
\usepackage{units}

\hypersetup{pdftitle={Search for photonic signatures of gauge-mediated supersymmetry in \unit[8]{TeV} $pp$ collisions with the ATLAS detector},pdfauthor={The ATLAS Collaboration}}

\AtlasTitle{Search for photonic signatures of gauge-mediated supersymmetry in \unit[8]{TeV} $pp$ collisions with the ATLAS detector}

\AtlasAbstract{%
A search is presented for photonic signatures motivated
by generalized models of gauge-mediated supersymmetry breaking. This
search makes use of \integLumi of proton-proton
collision data at $\sqrt{s} = \unit[8]{TeV}$ recorded by the ATLAS detector at the LHC, and explores
models dominated by both strong and electroweak production of supersymmetric
partner states. Four experimental signatures incorporating an isolated photon and
significant missing transverse momentum are explored. These signatures include events with
an additional photon, lepton, $b$-quark jet, or jet activity
not associated with any specific underlying quark flavor.
No significant excess of events is observed above the Standard Model prediction
and model-dependent 95\%~confidence-level exclusion limits
are set. 
}

\author{The ATLAS Collaboration}

\AtlasJournal{Phys.\ Rev.\ D}
\PreprintIdNumber{CERN-PH-EP-2015-168}
\date{17th July 2015}
\arXivId{}

\newcommand{\integLumi}{20.3{~\ifb}\xspace}
\newcommand{\integLumiPrior}{4.8{~\ifb}\xspace}
\newcommand{\integLumiE}{$20.3 \pm 0.6$ \ifb\xspace}

\newcommand{\GGMlimitG}{\unit[1290]{GeV}\xspace}

\newcommand{\GGMlimitW}{\unit[590]{GeV}\xspace}

\newcommand{\ptm}{\ensuremath{{\bf E}_{\mathrm{T}}^{\rm miss}}\xspace}
\newcommand{\dphi}{\ensuremath{\Delta \phi(\gamma,\MET)}\xspace}
\newcommand{\dphim}{\ensuremath{\Delta \phi_{\mathrm{min}}(\gamma,\MET)}\xspace}
\newcommand{\dphijm}{\ensuremath{\Delta \phi_{\mathrm{min}}(\mathrm{jet},\MET)}\xspace}
\newcommand{\dphijg}{\ensuremath{\Delta \phi_{\mathrm{min}}(\mathrm{jet},\gamma)}\xspace}
\newcommand{\HTj}{\ensuremath{H_{\mathrm{T}}^{\mathrm{jets}}}\xspace}
\newcommand{\RT}{\ensuremath{R_{\mathrm{T}}^{4}}\xspace}
\newcommand{\MTl}{\ensuremath{M_{\mathrm{T}}^{\ell,\MET}}\xspace}
\newcommand{\MTg}{\ensuremath{M_{\mathrm{T}}^{\gamma,\MET}}\xspace}
\newcommand{\MEFF}{\ensuremath{m_{\mathrm{eff}}}\xspace}

\newcommand{\Zgg}{\ensuremath{Z(\to \nu\bar{\nu})+\gamma\gamma}\xspace}
\newcommand{\Wgg}{\ensuremath{W(\to \ell\nu)+\gamma\gamma}\xspace}
\newcommand{\lgg}{\ensuremath{\ell \gamma\gamma}\xspace}
\newcommand{\QCDg}{\ensuremath{\mathrm{QCD}}\xspace}
\newcommand{\wlnu}{\ensuremath{W(\to \ell\nu)}\xspace}

\newcommand{\neutralino}{\ensuremath{\tilde{\chi}^{0}_{1}}\xspace}
\newcommand{\neutralinotwo}{\ensuremath{\tilde{\chi}^{0}_{2}}\xspace}
\newcommand{\neutralinothree}{\ensuremath{\tilde{\chi}^{0}_{3}}\xspace}
\newcommand{\chargino}{\ensuremath{\tilde{\chi}^{\pm}_{1}}\xspace}
\newcommand{\chiplus}{\ensuremath{\tilde{\chi}^{+}_{1}}\xspace}
\newcommand{\chiminus}{\ensuremath{\tilde{\chi}^{-}_{1}}\xspace}
\newcommand{\gravitino}{\ensuremath{\tilde{G}}\xspace}
\newcommand{\gluino}{\ensuremath{\tilde{g}}\xspace}

\newcommand{\wino}{\ensuremath{\tilde{W}}\xspace}
\newcommand{\winon}{\ensuremath{\tilde{W^0}}\xspace}

\newcommand{\mass}[1]{\ensuremath{m_{#1}}}

\newcommand{\Herwigpp}{{\tt HERWIG++}\xspace}
\newcommand{\Madgraph}{{\tt MADGRAPH}\xspace}
\newcommand{\Suspect}{{\tt SUSPECT}\xspace}
\newcommand{\Sdecay}{{\tt SDECAY}\xspace}

\newcommand{\Susyhit}{{\tt SUSY-HIT}\xspace}

\newcommand{\BS}{\ensuremath{\rm{SR}^{\gamma\gamma}_{S}}\xspace}
\newcommand{\BW}{\ensuremath{\rm{SR}^{\gamma\gamma}_{W}}\xspace}
\newcommand{\BSL}{\ensuremath{\rm{SR}^{\gamma\gamma}_{S-L}}\xspace}
\newcommand{\BSH}{\ensuremath{\rm{SR}^{\gamma\gamma}_{S-H}}\xspace}
\newcommand{\BWL}{\ensuremath{\rm{SR}^{\gamma\gamma}_{W-L}}\xspace}
\newcommand{\BWH}{\ensuremath{\rm{SR}^{\gamma\gamma}_{W-H}}\xspace}
\newcommand{\HBPL}{\ensuremath{\rm{SR}^{\gamma j}_{L}}\xspace}
\newcommand{\HBPH}{\ensuremath{\rm{SR}^{\gamma j}_{H}}\xspace}
\newcommand{\HBNL}{\ensuremath{\rm{SR}^{\gamma b}_{L}}\xspace}
\newcommand{\HBNH}{\ensuremath{\rm{SR}^{\gamma b}_{H}}\xspace}
\newcommand{\WNLSPe}{\ensuremath{\rm{SR}^{\gamma \ell}_{\it e}}\xspace}
\newcommand{\WNLSPu}{\ensuremath{\rm{SR}^{\gamma \ell}_{\mu}}\xspace}
\newcommand{\WNLSP}{\ensuremath{\rm{SR}^{\gamma \ell}_{{\it e}/\mu}}\xspace}

\begin{document}

\tableofcontents

\section{Introduction}
\label{sec:intro}
This paper reports on a search for four classes of events
containing energetic isolated photons and large missing transverse momentum
(with magnitude denoted \MET) in \integLumi of
proton-proton ($pp$) collision data at $\sqrt{s}=8$ TeV
recorded with the ATLAS detector at the Large Hadron Collider (LHC)
in 2012. For
the first of the four classes, two isolated energetic
photons are required (``diphoton'' events), while for the remaining classes
only a single isolated photon is required. For the second and third classes, the isolated photon
is required to appear in combination with
a ``$b$-jet'' identified as having arisen from the production of
a bottom ($b$) quark (``photon+$b$'' events) or an isolated electron or muon (``photon+$\ell$'' events), respectively. 
For the fourth class of events the isolated photon
is required to appear in combination with multiple jets selected
without regard to the flavor of the underlying parton (``photon+$j$'' events).

The results are interpreted in the context of a broad range of general models
of gauge-mediated supersymmetry breaking
(GGM)~\cite{Meade:2008wd,Buican:2008ws,Ruderman:2011vv}
that include the production of supersymmetric partners of
strongly coupled Standard Model (SM) particles as
well as SM partners possessing only electroweak charge.
In all models of GGM, the lightest supersymmetric particle (LSP)
is the gravitino \gravitino (the partner of the hypothetical quantum of the
gravitational field), with a mass significantly less than \unit[1]{GeV}.
In the GGM models considered here, the decay of the supersymmetric states produced in 
LHC collisions would proceed through the next-to-lightest supersymmetric particle
(NLSP), which would then decay to the \gravitino LSP and one or more SM particles,
with a high probability of decay into $\gamma$ + \gravitino.
In this study, several different possibilities for the nature of the NLSP
are considered, providing separate motivation for the four different and complementary 
experimental signatures that are
explored. In all models considered, all supersymmetric states with the exception of the \gravitino
are short lived, leading to prompt production of SM particles that are observed in the
ATLAS detector.

The results based on the diphoton and photon+$b$ signatures extend and supplant studies (Refs.~\cite{Aad:2012zza} and \cite{Aad:2012jva},
respectively) that made
use of \integLumiPrior of
$pp$ collision data at $\sqrt{s}=7$ TeV; the analyses based on the
photon+$j$ and photon+$\ell$ signatures are new and have only been performed with the 8 TeV data.
Making use
of 19.7~\ifb of $pp$ collision data at $\sqrt{s}=8$ TeV,
a search~\cite{CMS_photons_8Tev} for events similar in nature to those of the diphoton and photon+$j$ 
signatures mentioned above has performed by the CMS Collaboration,
and used to set limits on the masses of strongly coupled supersymmetric particles
in several GGM scenarios.

\section{Gauge-mediated supersymmetry phenomenology}
\label{sec:susy}
Supersymmetry~(SUSY)~\cite{Miyazawa:1966,Ramond:1971gb,Golfand:1971iw,Neveu:1971rx,Neveu:1971iv,Gervais:1971ji,Volkov:1973ix,Wess:1973kz,Wess:1974tw}
introduces a symmetry between fermions and bosons, resulting in a SUSY
partner (sparticle) with identical quantum numbers except a difference
by half a unit of spin for each SM particle. As none
of these sparticles have been observed, SUSY must be a broken symmetry
if realized in nature.  Assuming $R$-parity
conservation~\cite{Fayet:1976et,Fayet:1977yc,Farrar:1978xj,Fayet:1979sa,Dimopoulos:1981zb},
sparticles are
produced in pairs.  These would then decay through cascades involving
other sparticles until the stable, weakly interacting LSP is produced, leading 
to a final state with significant \MET.

Experimental signatures of gauge-mediated SUSY breaking
models~\cite{Dine:1981gu,AlvarezGaume:1981wy,Nappi:1982hm,Dine:1993yw, Dine:1994vc,Dine:1995ag}
are largely determined by the nature of the NLSP. 
For GGM, the NLSP is often formed from an admixture of any of the SUSY partners 
of the electroweak gauge and Higgs boson states.
In this study, three cases are assumed for the
composition of the NLSP. For the first case, the NLSP is assumed to be
purely binolike [the SUSY partner of the SM U(1) gauge boson]. For the
second case, the NLSP is assumed to be an admixture
of bino and neutral higgsino states.
For the final 
case, the NLSP is assumed to be a degenerate triplet
of wino states [the SUSY partners of the SM SU(2) gauge bosons].
In this paper, the neutral NLSP is denoted $\neutralino$ irrespective
of its composition. For the case that
the NLSP is a degenerate triplet, the charged NLSP states are denoted $\chargino$.
The properties of the GGM models used to represent these possibilities
are discussed below and summarized in Table~\ref{tab:GGM_models}.

For the case that the NLSP is a bino,
the final decay in each of the two cascades in a GGM event would be predominantly
$\neutralino\to\gamma+\gravitino$, leading
to final states with $\gamma\gamma+\met$.
For the case that the NLSP is a mixture of the bino
and higgsino, both the possibilities that the higgsino mass
parameter $\mu$ is less than or greater than zero are explored.
For the $\mu < 0$ possibility, the final decay in the cascade would include
a significant contribution from $\neutralino\to h + \gravitino$
with the subsequent decay $h \to \bbbar$, leading to final
states with a photon, multiple $b$-jets, and $\met$.
The latter ($\mu > 0$) possibility can produce scenarios for which the 
final decay in the cascade can be relatively evenly split 
between $\neutralino\to\gamma+\gravitino$ and
$\neutralino\to Z + \gravitino$, leading to final states with
a photon, multiple jets (including two from the hadronic decay of the $Z$ boson) 
that most often do not arise from $b$-quarks,
and $\met$.
For the case that the NLSP is a degenerate set of three wino states, 
the final step in the cascade includes charged as well as neutral wino decays.
Charged wino decays tend to produce isolated
leptons, while neutral wino decays produce photons with a wino-to-photon 
branching fraction that is no less than $\sin^2 \theta_W$ for any value of
the wino mass. Overall, these two wino-NLSP contributions lead to a significant
number of events with an isolated photon accompanied by an isolated lepton.
Of the five 
GGM models considered here, two (the ``gluino-bino'' and ``wino-bino'' models,
where the gluino is the SUSY partner of the gluon)
incorporate a purely binolike NLSP, two (the ``higgsino-bino'' models)   
incorporate a NLSP that is a higgsino-bino admixture, and one (the ``wino-NLSP'' model)
incorporates a winolike set of NLSPs; in all cases
the mass of the NLSP state is considered to be a free parameter
of the model.

\begin{table}[bp]
  \caption{Summary of the five GGM models considered in this study. For the two higgsino-bino
models, the functions $f_{\pm}(M_1,\mu)$ are chosen to establish NLSP decay properties
commensurate with the target experimental signature, as described in the text.}
  \begin{center}
  \begin{tabular*}{\textwidth}{@{\extracolsep{\fill}}lcccc} \hline\hline\noalign{\smallskip}
                            & Experimental       & Produced        & Composition    &  Free                \\
    {\bf GGM Model}         & Signature          & State(s)        & of NLSP        &  Parameters          \\
      \noalign{\smallskip}\hline\noalign{\smallskip}
    Gluino-bino             & diphoton           & gluino           & bino          & $M_{\gluino}$,$M_{\neutralino}$ \\   
    Wino-bino               & diphoton           & wino             & bino          & $M_{\wino}$,$M_{\neutralino}$   \\
    Higgsino-bino ($\mu<0$) & photon+$b$           & gluino, higgsino & higgsino/bino & $M_{\gluino}$,$f_{-}(M_1,\mu)$          \\
    Higgsino-bino ($\mu>0$) & photon+$j$           & gluino, higgsino & higgsino/bino & $M_{\gluino}$,$f_{+}(M_1,\mu)$          \\
    Wino NLSP               & photon+$\ell$      & wino             & wino          & $M_{\wino}$          \\
      \hline\hline
  \end{tabular*}
  \label{tab:GGM_models}
  \end{center}
\end{table}

The two 
GGM models incorporating a binolike NLSP
are the focus of the diphoton analysis. For these models,
one other set of SUSY partner states is taken to be potentially 
accessible in 8 TeV {\it pp} collisions, while
all other SUSY masses are decoupled (set to
inaccessibly large values). For both 
of these binolike NLSP cases, production
proceeds solely through this set of SUSY partners,
with the NLSP appearing in the subsequent
decays of the produced SUSY partner states. For the gluino-bino
model, the set of partners is composed of a degenerate octet of
gluinos.
For the wino-bino model, the set of partners is composed
of a degenerate triplet of wino states
\neutralinotwo, \chargino, and is dominated by the production
of \chiplus\chiminus and \neutralinotwo\chargino.
For both
of these models, the masses of these produced states
are considered to be free parameters along with that of the chosen
\neutralino state, the latter of which is constrained to be less
than those of the produced states.
This results in a SUSY production
process that proceeds through the creation of pairs of the higher-mass
states, which subsequently decay through
short cascades to the NLSP \neutralino states. Other SM
objects (jets, leptons, photons) may be produced in these cascades.
The \neutralino branching fraction
to $\gamma$ + \gravitino is 100\% for $m_{\neutralino} \to 0$ and
approaches $\cos^2 \theta_W$ for $m_{\neutralino} \gg m_Z$, with
the remainder of the \neutralino sample decaying to $Z$ + \gravitino.
For all \neutralino masses, then, the branching fraction is dominated
by the photonic decay, leading to the diphoton-plus-\met signature.
For these models with a binolike NLSP, typical production and decay channels for strong (gluino) and
electroweak (wino) production are exhibited in Fig.~\ref{fig:feynman_bino}.

\begin{figure}[tp]
  \begin{center}
    \includegraphics[width=0.45\textwidth]{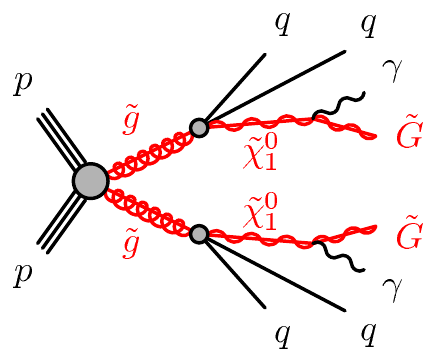} ~~
    \includegraphics[width=0.45\textwidth]{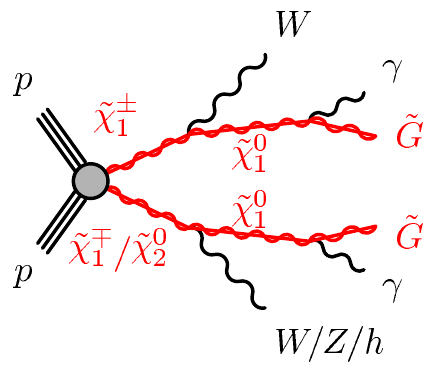}
  \end{center}
  \caption{
Typical production and decay-chain processes for the gluino-production (left)
      and electroweak-production (right) instances of the 
      GGM model for which the NLSP is a binolike neutralino,
referred to in the text as the gluino-bino and wino-bino models, respectively.
    \label{fig:feynman_bino}
  }
\end{figure}

\begin{figure}[tp]
  \begin{center}
    \includegraphics[width=0.45\textwidth]{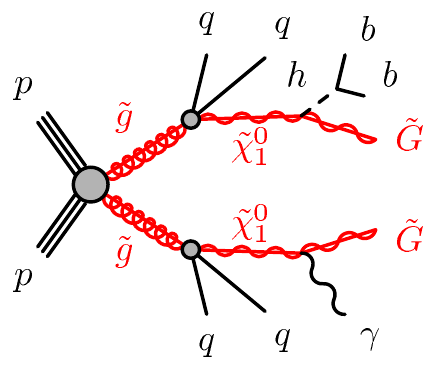} ~~
    \includegraphics[width=0.45\textwidth]{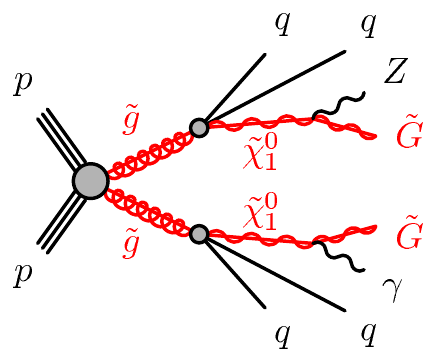}
  \end{center}
  \caption{
    Typical production and decay-chain processes for the gluino-production
      instance of the
      GGM model for which the NLSP is a higgsino-bino neutralino admixture,
      referred to in the text as the higgsino-bino model. 
      For the model with $\mu < 0$ (left), the final
      step of the cascade (the \neutralino decay) would have a
      probability of order 50\% of producing a Higgs boson rather than a
      photon or $Z$ boson; for the model
      with $\mu > 0$ (right), the \neutralino decay would have a
      probability of order 50\% of producing a $Z$ boson rather than a photon.
      For both of these models, production can also proceed
      through gaugino and neutralino states, which can dominate the production
      cross section for high values of gluino mass.
    \label{fig:feynman_higgsino_bino}
  }
\end{figure}

\begin{figure}[tp]
  \begin{center}
    \includegraphics[width=0.45\textwidth]{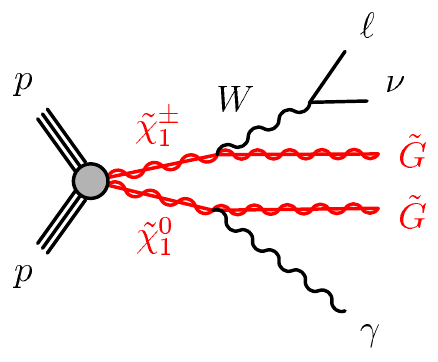}
  \end{center}
  \caption{
Typical production and decay-chain processes for the wino-NLSP model.
In this model, the \neutralino is a pure \winon state, while the
\chargino are the two charged wino states.
    \label{fig:feynman_wino}
  }
\end{figure}

The higgsino-bino GGM models incorporate a NLSP composed of
a higgsino-bino admixture, as well as
a degenerate octet of gluinos 
identical in nature to those of the
gluino-bino model. 
For the first 
of these models, which is the focus of the photon+$b$ analysis,
the higgsino mass parameter $\mu$ 
is required to be negative, and the composition
of the NLSP is set by adjusting $\mu$ and the GGM U(1) mass parameter
$M_1$ so that a constant
ratio of the branching fraction of $\neutralino \to h + \gravitino$
to that of $\neutralino \to \gamma + \gravitino$ is maintained at
approximately 1.7:1 over the full range of NLSP masses.
The photon+$b$ analysis
was found to provide the greatest advantage relative to the
diphoton analysis for this ratio of branching fractions.
In the limit that $m_{\neutralino} \gg m_Z$, the NLSP
branching fractions to $h + \gravitino$, $\gamma + \gravitino$,
and $Z + \gravitino$ approach 56\%, 33\%, and 11\%, respectively.
The GGM SU(3) mass parameter $M_3$ bears a direct relation to the
gluino mass, and is taken to be a free parameter in this $\mu < 0$ higgsino-bino
model, with all squark states decoupled. 
The GGM SU(2) mass parameter $M_2$ is set to a value of 2.5 TeV.
Four other electroweak gaugino states typically lie
within 25 GeV of the \neutralino NLSP: the two lightest charginos \chargino, and two
additional neutralinos
\neutralinotwo and \neutralinothree. 
The pair production of gluinos or any of these four additional gaugino
states leads to decays to the \neutralino via
cascades involving SM particles.

For the second of the higgsino-bino models, which is
the focus of the photon+$j$ analysis, the $\mu$ parameter is 
chosen to be positive, which suppresses the $h + \gravitino$
decay mode of the higgsino. As in the models described above,
the NLSP mass is taken to be a
free parameter.
The $M_1$ and $\mu$ parameters are adjusted so that the branching
fractions of the $\neutralino$ to $\gamma + \gravitino$, $Z + \gravitino$
and $h + \gravitino$ are maintained close to 50\%, 49\% and 1\% for
most values of the $\neutralino$ and gluino masses.
In this model, the production of gluino pairs can be followed
by decays to both a single photon and a hadronically decaying $Z$ boson, producing 
events with a single isolated high-energy photon accompanied 
by two jets.
In the case that the gluino mass is substantially
larger than the $\neutralino$ mass, additional jets can
be produced in the cascade.
Three additional electroweak gaugino states
lie close in mass to the \neutralino, allowing for the
possibility of SUSY production through pairs of these
states. Such events tend to produce
fewer jets than those that proceed through gluino production,
but in certain regions of the model space can provide a significant
contribution to data samples selected to isolate the photon-plus-jets
signature. As in the $\mu < 0$ higgsino-bino model, the value of $M_3$, which is directly
related to the gluino mass, is taken to be a free parameter, $M_2$ is
set to a value of 2.5 TeV, and all squark states are decoupled.
Typical production and decay-chain processes for the two models
for which the NLSP is a higgsino-bino admixture are shown in Fig.~\ref{fig:feynman_higgsino_bino}.

Finally, the wino-NLSP model, which is the focus of the photon+$\ell$ analysis,
incorporates a set of three degenerate winolike
NLSPs. This set includes the neutral \winon, which as the lightest
neutral gaugino is also referred to as the \neutralino,
as well as the two charged wino states, which form the \chargino states.
Production proceeds 
through the direct production of pairs of NLSP states; 
such events usually contain at least one \winon NLSP. Although
the \winon couples preferentially to the $Z$ boson relative to
the photon, the \winon decays
into a photon+gravitino final state with unit branching fraction
for wino mass below that of the $Z$ boson.
The \winon branching fraction to photon+gravitino approaches
$\sin^2 \theta_W$ for wino masses far above that of the $Z$ boson.
Leptons can be produced either through the decays of charged wino
states, or through the decays of $Z$ bosons that arise from \winon
decay, leading to a significant probability that the overall final state
would contain both a photon and a lepton. In this model, 
a common wino mass scale is taken as a free parameter, with
all other GGM mass parameters
set to a value of \unit[2.5]{TeV}, except the squark masses,
which are set to infinity.
A production and decay diagram typical for this model is shown in Fig.~\ref{fig:feynman_wino}.

For all five
models considered here, the mass of the gravitino is chosen so that
the NLSP decay length is never greater than 1 mm. This ensures that all
particles arising from the decay of the NLSP are prompt, and in particular
that the relationship between the point and direction of impact
of photons from NLSP decay upon the face of the detector is
consistent with that of a prompt photon (a separate analysis~\cite{Aad:2014gfa}
searches for GGM models with a longer-lived binolike NLSP, leading to signatures with nonprompt photons).
In addition, the ratio 
$\tan \beta$ of the two SUSY Higgs-doublet vacuum-expectation values is set to a value of 1.5;
for all five
models, the
phenomenology relevant to this search is only
weakly dependent on the value of $\tan \beta$.

\section{Samples of simulated processes}
\label{sec:MC}

For the GGM models under study, the SUSY mass spectra and branching ratios are calculated using
\Suspect 2.41~\cite{Djouadi:2002ze} and
\Sdecay~1.3b~\cite{Muhlleitner:2003vg}, respectively, inside the package \Susyhit 1.3~\cite{Djouadi:2006bz}.
The Monte Carlo~(MC) SUSY signal samples are produced using
\Herwigpp~2.5.2~\cite{Bahr:2008pv} with {\tt CTEQ6L1} parton distribution
functions~(PDFs)~\cite{Pumplin:2002vw}.
Signal cross sections are calculated to next-to-leading order~(NLO) in the strong coupling constant,
including, for the case of strong production, the resummation of soft gluon emission at next-to-leading-logarithmic accuracy
(NLO+NLL)~\cite{Beenakker:1996ch,Kulesza:2008jb,Kulesza:2009kq,Beenakker:2009ha,Beenakker:2011fu}.
The nominal cross section and its uncertainty are taken from an envelope of cross-section predictions
using different PDF sets and factorization and renormalization scales~\cite{Kramer:2012bx}. 
At fixed center-of-mass energy, SUSY production cross sections decrease rapidly with increasing SUSY partner mass.
At $\sqrt{s} = 8$ TeV, the gluino-production cross section is approximately \unit[24]{fb}
for a gluino mass of \unit[1000]{GeV} and falls to below \unit[1]{fb} for a gluino
mass of \unit[1400]{GeV}. The wino-production cross section is approximately \unit[15]{fb} for a wino mass
of \unit[500]{GeV}, and falls to approximately \unit[1]{fb}
for a wino mass of \unit[750]{GeV}.

While most of the backgrounds to the GGM models under examination
are estimated through the use of control samples selected from data, as described below,
the extrapolation from control regions (CRs) to signal regions (SRs) depends on simulated samples, as do the optimization studies.
The simulation of $W$ and $Z$ boson production, including events with up to five accompanying partons, is calculated by two different generators.
The {\tt ALPGEN} 2.14~\cite{Mangano:2002ea}
Monte Carlo generator is interfaced to {\tt HERWIG}
6.520 for showering and fragmentation and to
{\tt JIMMY}~\cite{Butterworth:1996zw} for simulation of the underlying event.
Parton distributions are provided by the {\tt CTEQ6L1} functions.
Similar samples are produced with the {\tt SHERPA} 1.4.1 generator~\cite{sherpa} with
{\tt CT10}~\cite{Lai:2010vv} PDFs, for up to four accompanying partons.

$W\gamma$ production
is also simulated via {\tt ALPGEN} interfaced
to {\tt HERWIG} and {\tt JIMMY}, but makes use of the {\tt CT10} PDFs.
Other $W\gamma$ samples are generated, as is
the $Z\gamma$ process, by using {\tt SHERPA}
with the {\tt CT10} PDFs.
The $\ttbar\gamma$ process is simulated at leading order (LO) using {\Madgraph} 5.1.5.11~\cite{Alwall:2007st} and {\tt CTEQ6L1},
interfaced to the {\tt PYTHIA} 6.427 parton
shower generator~\cite{Sjostrand:2006za}.
The \ttbar\ process is simulated not only with the
the {\tt POWHEG} generator interfaced to {\tt PYTHIA} and the {\tt CTEQ6L1} PDFs,
but also with the {\tt MC@NLO} 4.06 generator~\cite{Frixione:2002ik,mcatnlo2} and the {\tt CT10} PDFs, including full
NLO QCD corrections.
This contribution is rescaled to match the \ttbar cross section 
at NNLO with NNLL soft gluon terms, as calculated 
with top++2.0~\cite{Cacciari:2011hy,Baernreuther:2012ws,Czakon:2012zr,Czakon:2012pz,Czakon:2013goa,Czakon:2011xx}.
The $t\gamma$ and $\bar{t}\gamma$ processes are simulated with the {\tt WHIZARD
  2.1.1}~\cite{whizard, whizard2}
generator, with four-flavor/five-flavor matching provided using {\tt HOPPET}~\cite{Salam:2008qg}.
Additional photon radiation is added with {\tt PHOTOS}~\cite{photos}, with 
parton showering and fragmentation again simulated with {\tt PYTHIA}.
Other $t$ and $\bar{t}$ samples are generated with {\tt POWHEG}.

The $\gamma$+jet(s) process is simulated in a similar
manner to the $W^\pm$ or $Z$ samples using {\tt ALPGEN} interfaced to
{\tt HERWIG} and {\tt JIMMY} and the {\tt CTEQ6L1} PDFs.
A generator-level requirement of $\unit[35]{GeV}$ is applied
to the photon transverse momentum $p^{\gamma}_{\rm T}$, and
the sample is generated in exclusive bins of $p^{\gamma}_{\rm T}$ to produce a more
statistically significant sample at higher values of  $p^{\gamma}_{\rm T}$.
Additional $\gamma$+jet(s) samples are used,
simulated with {\tt SHERPA} and the {\tt CT10} PDFs.
The prompt diphoton sample is generated with {\tt PYTHIA} 6.423, which
includes the subprocesses
$gg \rightarrow \gamma\gamma$ and $q\bar{q}
\rightarrow \gamma\gamma$, with the
requirement that there be at least two prompt photons with generated 
transverse momentum greater than 20 GeV.
Parton densities are modeled according to the
{\tt MRST\,2007\,LO}${}^{*}$~\cite{Sherstnev:2007nd} functions.

The background from \Zgg production is simulated
using the {\tt SHERPA} MC generator,
normalized to a cross section calculated at LO using \Madgraph~5~ and the
{\tt CTEQ6L1} PDF, and then corrected by a $K$-factor of $2.0 \pm 1.0$~\cite{Bozzi:2011en}.
The background from \Wgg production is simulated using the
{\tt ALPGEN} MC generator,
although the overall normalization is set via a study making use of data events containing two photons
and a charged lepton (to be discussed below).
Diboson production, for the case that each boson is a $W$ or $Z$, is simulated with {\tt POWHEG}.

All MC samples are processed with the {\tt GEANT4}-based
simulation~\cite{Agostinelli:2002hh} of the ATLAS
detector~\cite{Aad:2010ah}, or, where appropriate, a simulation of the
ATLAS detector based on parametrized shower shapes in the
calorimeter, and {\tt GEANT4} elsewhere.  Corrections are applied to
the simulated samples to account for differences between data and
simulation for the lepton and photon trigger, identification, and
reconstruction efficiencies, as well as for the efficiency and
misidentification rate of the algorithm used to identify jets
containing $b$-hadrons ($b$-tagging).  The variation of the number of
$pp$ interactions per bunch crossing (``pileup'') as a function of the
instantaneous luminosity is taken into account by overlaying simulated
minimum-bias events according to the observed distribution of the
number of pileup interactions in data, with an average of $21$
interactions per event.

\newcommand{\AtlasCoordFootnote}{%
ATLAS uses a right-handed coordinate system with its origin at the nominal interaction point (IP)
in the center of the detector and the $z$ axis along the beam pipe.
The $x$ axis points from the IP to the center of the LHC ring,
and the $y$ axis points upwards.
Cylindrical coordinates $(r,\phi)$ are used in the transverse plane, 
$\phi$ being the azimuthal angle around the beam pipe.
The pseudorapidity is defined in terms of the polar angle $\theta$ as $\eta = -\ln [\tan(\theta/2)]$.
Angular distance is measured in units of $\Delta R \equiv \sqrt{(\Delta\eta)^2 + (\Delta\phi)^2}$.}

\section{ATLAS detector}
The ATLAS experiment makes use of a multipurpose detector~\cite{Aad:2008zzm}
with a forward-backward symmetric cylindrical geometry and nearly
4$\pi$ solid angle coverage.\footnote{\AtlasCoordFootnote}
Closest to the beam line are solid-state tracking
devices comprising layers of silicon-based pixel and strip detectors
covering $\left|\eta\right|<2.5$ and straw-tube
detectors covering $\left|\eta\right|<2.0$, located inside a thin
superconducting solenoid that provides a \unit[2]{T} magnetic field.
Outside the solenoid, fine-grained lead/liquid-argon
electromagnetic (EM) calorimeters provide coverage over
$\left|\eta\right| < 3.2$ for the measurement of the energy and position of
electrons and photons.
A presampler, covering $\left|\eta\right| < 1.8$, is used to correct
for energy lost upstream of the EM calorimeter.  An
iron/scintillator-tile hadronic calorimeter covers the region $|\eta|
< 1.7$, while a copper/liquid-argon medium is used for hadronic
calorimeters in the end cap region $1.5 < |\eta| < 3.2$. In the
forward region $3.2 < |\eta| < 4.9$ liquid-argon calorimeters with
copper and tungsten absorbers measure the electromagnetic and hadronic
energy.  A muon spectrometer consisting of three superconducting
toroidal magnet systems, each comprising eight toroidal coils,
tracking chambers, and detectors for
triggering, surrounds the calorimeter system.
The muon system reconstructs penetrating tracks over a range $|\eta| < 2.7$
and provides input to the trigger system over a range $|\eta| < 2.4$.
A three-level trigger system is used to select events.
The first-level trigger is implemented in hardware and uses a subset of the detector information
to reduce the accepted rate to at most 75 kHz.
This is followed by two software-based trigger levels that
together reduce the accepted event rate to 400 Hz on average
depending on the data-taking conditions during 2012.

\section{Reconstruction of candidates and observables}
\label{sec:objects}

Primary vertices are formed from sets of two or more tracks, 
each with transverse momentum $p_{\rm T}^{\mathrm{track}} >$ \unit[400]{MeV}, that are
mutually consistent with having originated at the same three-dimensional
space point within the luminous region of the colliding proton beams.
When more than one such
primary vertex is found, the vertex with the largest scalar sum
of the squared transverse momenta of the associated tracks is chosen.
To further ensure the event resulted from
a beam collision, the primary vertex of the event is required to have at 
least five associated tracks.

Electron
candidates are reconstructed from EM calorimeter energy clusters
consistent with having arisen from the impact of an electromagnetic
particle (electron or photon) upon the face of the calorimeter.
For the object to be considered an electron, it is required to
match a track identified by a reconstruction algorithm 
optimized for recognizing charged particles with a high probability of bremsstrahlung. 
In addition, the matched track
is required to include information from at least seven layers of the solid-state tracking
system; a track within the acceptance of the tracking system
typically traverses eleven layers of the solid-state tracking system.
The energy of the electron candidate
is determined from the EM cluster, while its pseudorapidity is
determined from the associated reconstructed track. 
Further details of the reconstruction of electrons can be found in
Refs.~\cite{Aad:2014fxa} and \cite{ATLAS-CONF-2014-032}.
Electron
candidates used by these analyses are further required to have
$\pt > \unit[20]{GeV}$ and $|\eta| < 2.47$. For the photon+$\ell$
analysis, signal electrons are not allowed to be within the
transition region $1.37 < \left|\eta\right| < 1.52$ between the barrel
and end cap calorimeters. A track-based isolation requirement is
imposed, with the scalar sum of the transverse momenta of tracks within a cone
of size $\Delta R = 0.3$ required to be less than 16\% of the
electron \pt. Finally, the electron track is required to be consistent with
coming from the primary vertex in the $r$--$z$ plane.

Electromagnetic clusters are classified as photon candidates provided
that they either have no matched track 
or have
one or more matched tracks consistent with coming from a photon conversion vertex.
Based on the 
characteristics of the longitudinal and transverse shower development
in the EM calorimeter, photons are classified as
``loose'' or ``tight.'' 
Further details of the reconstruction of photons can be found in
Ref.~\cite{Aad:2010sp}.
In the case that an EM calorimeter
deposition is identified as both a photon and an electron, the photon
candidate is discarded and the electron candidate retained.
Photon candidates used by these analyses are required to be within
$\left|\eta\right| < 2.37$, and to be outside the transition region
$1.37 < \left|\eta\right| < 1.52$. 
Finally, an isolation requirement is imposed.  After correcting for
contributions from pileup and the deposition ascribed to the photon
itself, loose and tight isolation criteria are defined, with the
tight criterion requiring less than \unit[4]{GeV} of transverse
``isolation energy'' in a cone of size $\Delta R = 0.4$ surrounding the
energy deposition in the calorimeter associated with the photon.  For
the loose isolation criterion, no more than \unit[5]{GeV} of isolation
energy is allowed within a cone of size $\Delta R = 0.2$.  The tight
criterion is used for the diphoton analysis, while the loose
criterion is used for the remaining three signatures (photon+$b$,
photon+$j$, photon+$\ell$).

Muon candidates make use of reconstructed tracks from the tracking system 
as well as information from the muon system~\cite{Aad:2014rra}. 
Muons are required to be either
``combined,'' for which the muon is reconstructed independently in both the
muon spectrometer and the tracking system and then combined, or ``segment-tagged,'' for which
the muon spectrometer is used to tag tracks as muons,
without requiring a fully reconstructed candidate in the muon spectrometer.
Signal muons are required to have $\pt > \unit[20]{GeV}$
and $|\eta|<2.4$.  Track-based as well as calorimeter-based isolation
requirements are imposed, with the scalar sum of the transverse momenta of
tracks within a cone of size $\Delta R = 0.3$ required to be less than
12\% of the muon \pt, and the energy in the calorimeter projected in
the transverse plane within a cone of size $\Delta R = 0.3$, corrected for pileup, also required
to be less than 12\% of the muon \pt. Finally, the muon track
is required to be consistent with coming from the primary vertex in both the
$r$--$z$ and $r$--$\phi$ planes.

Jets are reconstructed from three-dimensional calorimeter energy clusters using the anti-$k_t$ 
algorithm~\cite{Cacciari:2008gp} with a
radius parameter $R$ = 0.4. Jets arising from detector noise, cosmic rays or other noncollision sources are rejected,
as described in Ref.~\cite{Aad:2011he}. 
Each cluster is classified, prior to the jet reconstruction, as coming from an 
electromagnetic or hadronic
shower on the basis of its shape~\cite{clustering}. Each cluster energy is then corrected by weighting electromagnetic
and hadronic energy deposits with correction factors derived from Monte Carlo simulation. A correction is applied to
subtract the expected contamination from pileup, calculated as the product of the jet area in $\eta$--$\phi$ space
and the average energy density of the event~\cite{Aad:2014bia}. A further calibration, relating the response of the calorimeter to
\emph{in situ} jet-energy measurements~\cite{Aad:2014bia} is then applied.
Once calibrated, jets are required to have $\pt >$ \unit[20]{GeV} and $|\eta| < 2.8$.
Jets containing $b$-hadrons are identified using
the MV1c $b$-tagging algorithm~\cite{ATLAS-CONF-2014-046}.
This neural network algorithm combines the information from
various algorithms based on track impact-parameter significance or
explicit reconstruction of $b$- and $c$-hadron decay vertices.
The analyses presented in this paper use an operating point corresponding to 70\% efficiency
for jets originating from the fragmentation of a $b$-quark in simulated
\ttbar events, selecting approximately 0.7\% of light-quark
and gluon-induced jets 
and 20\% of $c$-quark-induced jets.

In the case that two reconstructed objects are in close enough proximity to one 
another to raise a concern that they are a single detector object 
reconstructed as more than one particle or jet candidate, an overlap-removal
procedure is followed. 
To reduce the rate of electrons misidentified as photons, if the angular distance
$\Delta R$ between a reconstructed electron and photon is less than 0.01, 
the object is classified as an electron.
 
To avoid ambiguity that arises when an electron or photon is also reconstructed as a jet,
if a jet and an electron or photon are reconstructed within an angular distance
$\Delta R = 0.2$ of one another, the electron or photon is retained and
the jet is discarded; if $0.2 < \Delta R < 0.4$
then the jet is retained and the electron or photon is discarded.
Finally, in order to suppress the reconstruction of muons arising from showers induced 
by jets, if a jet and a muon are found with $\Delta R < 0.4$ the jet is retained 
and the muon is discarded.

The vector momentum imbalance in the transverse plane is obtained from the negative vector
sum of the reconstructed and calibrated physics objects and is referred to as missing transverse
momentum \ptm~\cite{ref:MET}.
Calorimeter energy deposits
are associated with a reconstructed and identified high-$\pt$ object in a specific order: electrons with
$\pt > 10$ GeV, photons with $\pt > 10$ GeV, and jets with $\pt > 20$ GeV.
Deposits not associated with any such objects are also taken into account in the \ptm 
determination, as are muons with $\pt > 10$ GeV. 

The transverse mass $M_T$ of a system of two massless particles with
four-vectors $p_1$ and $p_2$ is
given by
$$M_{\rm T} = \sqrt{2 p_{{\rm T},1} p_{{\rm T},2} (1 - \cos \Delta\phi_{1,2})}, $$
where $\Delta\phi_{1,2}$ is the angular separation between the 
two vectors projected into the transverse plane.
The analyses presented here make use of the transverse mass of both the 
photon-\met ($\MTg$) and 
lepton-\met ($\MTl$) systems, where the lepton is taken to be massless in 
the transverse-mass determination.

Several additional observables are defined to help in the discrimination of SM backgrounds
from potential GGM signals. The total visible transverse energy
\HT is calculated as the scalar sum of
the transverse momenta of the selected photons and any
additional leptons and jets in the event; a similar observable based only
on the momenta of jets in the events is referred to as \HTj. The ``effective mass'' \MEFF
is defined as the scalar sum of \HT and \met.
The photon-\met separation
\dphi is defined as the azimuthal
angle between the missing transverse momentum vector and the 
selected photon. In the case of the diphoton analysis, \dphim is 
defined to be the minimum value of \dphi of the two
selected photons. The minimum jet-\met separation \dphijm
is defined as the minimum azimuthal angle between
the missing transverse momentum vector and the leading (highest-\pt)
jets in the event. 
The number of leading jets used differs depending on the signature
under study and is shown in Tables~\ref{tab:dp-signalregion} and~\ref{tab:hb-signalregion}.
For the diphoton analysis, leading jets are required
to have $\pt >$ \unit[75]{GeV}, and if no such jet is found, no requirement is placed on 
the observable. 
The quantity 
\dphijg is defined as the minimum separation between the selected photon
and each of the two leading jets in the event. 
The quantity $\Delta R(\ell,\gamma)$ is defined as the distance in $\eta$--$\phi$
space between the leading photon and lepton.
Finally, the quantity \RT is defined as
the scalar sum of the transverse momentum of the four highest-\pt
jets in the event divided by the sum of the
transverse momentum of all jets in the event.

\section{Event selection}
\label{sec:selection}

The data sample
is selected by a trigger
requiring the presence of one loose photon
with energy projected into the plane transverse to the beam pipe ($\ET$) of greater than 120 GeV for the
photon+$b$, photon+$j$ and photon+$\ell$ analyses, or two loose photons
with $\ET > 40$ GeV for the diphoton analysis.
Events are removed from the data sample if they 
contain jets likely to be produced by beam backgrounds, cosmic rays
or detector noise, as described in Ref.~\cite{Aad:2011he}.
After applying data-quality requirements related to the beam and detector conditions, the total
available integrated luminosity is \integLumi. The uncertainty on the integrated luminosity is $\pm2.8\%$,
estimated via the methodology of Ref.~\cite{lumi2011}.

For the diphoton analysis, geared towards the exploration of 
the gluino-bino and wino-bino GGM models incorporating a purely binolike \neutralino,
two separate SR selection strategies
were developed: a ``\BS'' selection geared towards the production
of higher-mass strongly coupled SUSY states (gluinos and squarks) and a
``\BW'' selection geared towards the production of lower-mass
weakly coupled SUSY states (winos). For each of these approaches,
two SRs are defined: the first (\BSL, \BWL) optimized for the
case of a lower-mass \neutralino and the second (\BSH, \BWH) for
a higher-mass \neutralino. 

For the photon+$b$ analysis, geared towards the higgsino-bino
GGM model with a negative value of the $\mu$ parameter, 
two SRs (\HBNL, \HBNH) are defined. The SRs are again distinguished
by their optimization for low  and high \neutralino mass,
respectively. In particular, the \HBNL selection is designed to
have a high acceptance for events that arise through the production
of pairs of weakly coupled SUSY partners, which can have a significant
cross section for the low-\neutralino-mass reaches of the
higgsino-bino GGM model explored here.
For the photon+$j$ analysis, geared towards the higgsino-bino
GGM model with a positive value of the $\mu$ parameter,
a further set of two SRs are defined (\HBPL, \HBPH).
These two SRs are once again distinguished
by their optimization for low and high \neutralino mass, respectively.

A final ``\WNLSP'' signal region was developed to search for photon+$\ell$ events arising
from the GGM model with a winolike set of NLSPs. This SR is divided
into two subsets---\WNLSPe and \WNLSPu---depending on the flavor of
the leading lepton (electron or muon).

All four diphoton SRs require two tight, isolated photons with $\ET > \unit[75]{GeV}$,
while the \HBNL and \HBNH signal regions require a single tight, isolated photon with $\ET > \unit[125]{GeV}$
and $\ET > \unit[150]{GeV}$, respectively, and the 
\HBPL and \HBPH signal regions require a single tight, isolated photon with $\ET > \unit[125]{GeV}$
and $\ET > \unit[300]{GeV}$, respectively. The \WNLSP signal region requires a single tight, isolated
photon with $\ET > \unit[125]{GeV}$.
Along with \met, leptonic, and ($b$-)jet
activity, requirements are
made on a number of additional observables, with values chosen to optimize
the sensitivity to the GGM signal of interest for each SR.
To ensure that the \met observable is accurately measured, minimum requirements
on \dphim and \dphijm are considered for each SR. 
For the \HBPH signal region of the photon+$j$ analysis, rejecting events with 
jets misidentified as photons by placing a 
requirement on \dphijg is found to improve the sensitivity of the analysis.

To exploit the
high-energy scale associated with SUSY production at masses close to
the expected limit of sensitivity of the various SRs, several SRs include minimum requirements
on one of the two total-transverse-energy observables \HT or \MEFF.
As an illustration,
Fig.~\ref{fig:diphoton_meff}~(left) shows the \MEFF distribution of selected diphoton
events as well as that expected from several SM sources and from 
characteristic strong-production points of the binolike NLSP GGM model.
For electrons from $W$ boson decay that are misreconstructed as photons, the transverse mass \MTg of the
photon-\ptm system tends to be less than that of the $W$ boson; because of this, the photon+$b$
analysis is found to benefit from a minimum requirement on \MTg. The \HBNL analysis also benefits from a requirement that the 
invariant mass $M_{bb}$
of the system formed by the two most energetic $b$-jets be close to the Higgs boson mass.
A minimum requirement on the transverse mass \MTl of the lepton-\ptm system
is similarly found to be effective in rejecting backgrounds from $W$ boson 
and semileptonic \ttbar decay for the photon+$\ell$ analysis. A further requirement that the 
electron-photon system invariant
mass not be 
close to the $Z$ boson mass helps to reject
$Z$ boson backgrounds to the photon+$\ell$ analysis.
A requirement that \HBNL signal events have
no identified charged leptons helps to reduce the background from
semileptonic \ttbar events, while a requirement that \HBPH signal
events have $\RT < 0.85$ helps reduce the background from SM events, 
which tend to have fewer and softer jets than do signal events; as an illustration,
see Fig.~\ref{fig:diphoton_meff}~(right).
Finally, a requirement that the total transverse energy from jets
with $\pt > \unit[40]{GeV}$ (\HTj) be
less than \unit[100]{GeV} helps reduce the backgrounds to \WNLSP due to 
top quark production.

\begin{figure}[tp]
  \begin{center}
    \includegraphics[width=0.48\textwidth]{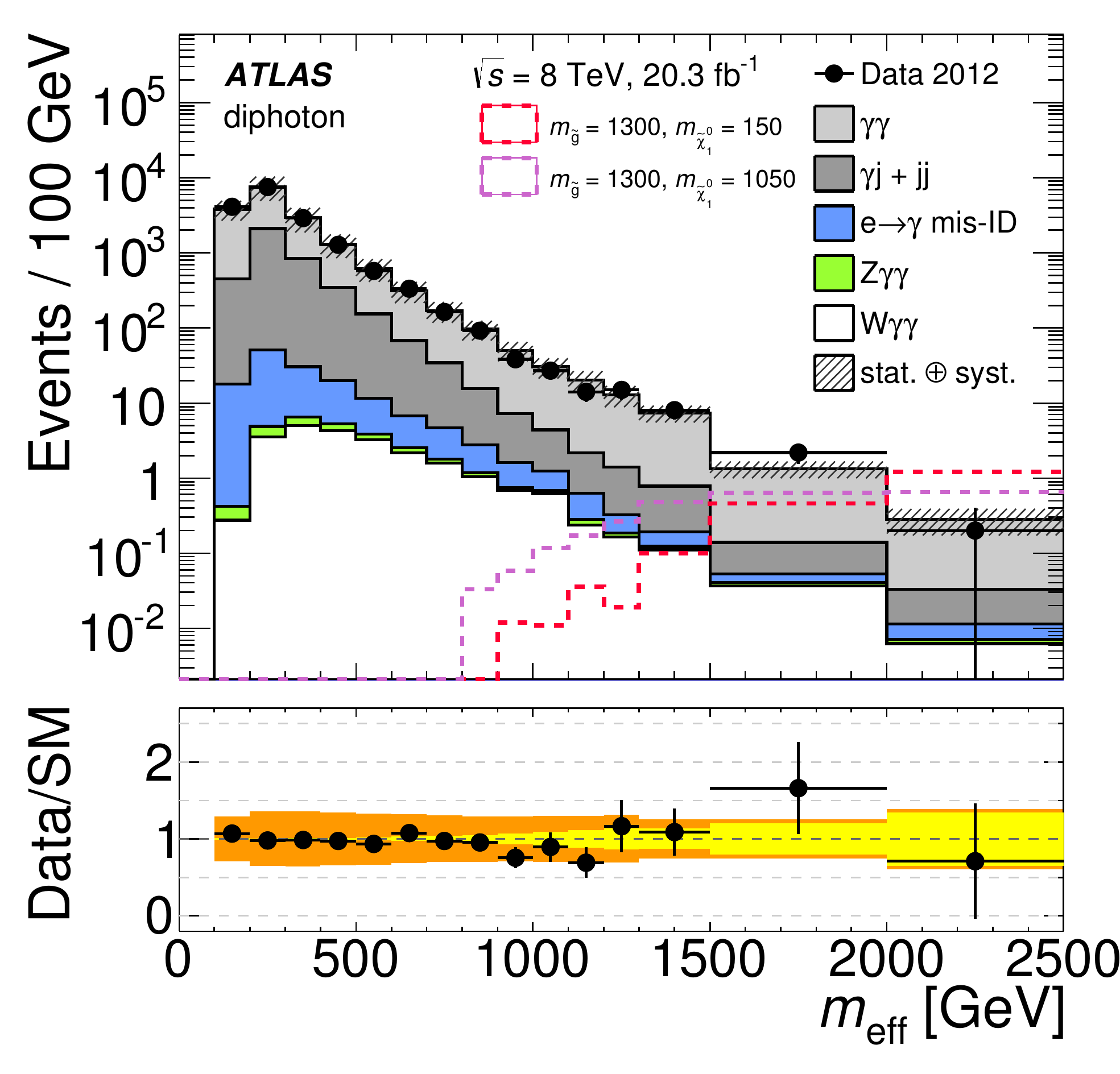} ~~
    \includegraphics[width=0.48\textwidth]{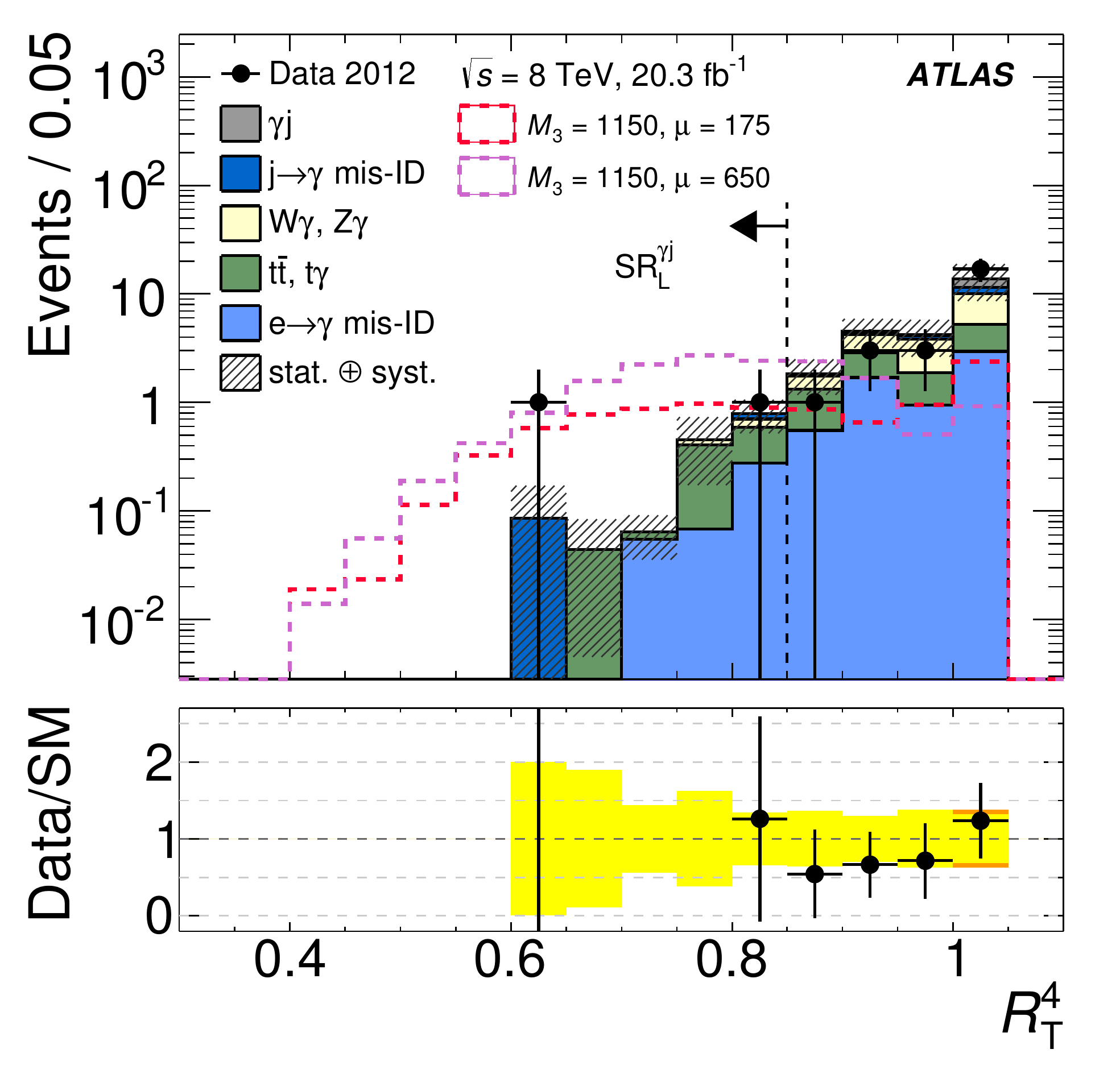}
  \end{center}
  \caption{
    (Left)~Distribution of \MEFF, the sum of the total visible transverse energy and \met,
    for selected diphoton events, after requiring $\dphijm > 0.5$ but before 
    application of a requirement on \MET and \dphim. Also shown are the
    expected contributions of SM processes, estimated as 
    described in Sec.~\ref{sec:backgrounds}, as well as
    the expected \MEFF distributions 
    for the $(\mass{\gluino},\mass{\neutralino}) = (1300,150)$ GeV
    and $(\mass{\gluino},\mass{\neutralino}) = (1300,1050)$ GeV gluino-bino GGM models.
    (Right)~Distribution of \RT, the scalar sum of the transverse momentum of
    the four highest-$p_T$ jets in the event divided by the sum of the transverse
    momentum of all jets in the event, for the sample surviving all \HBPL selection
    criteria except the \RT requirement itself. 
    Also shown are the
    expected contributions of SM processes, estimated as
    described in Sec.~\ref{sec:backgrounds}, as well as
    the signal expectation
    for the two points in the $M_3$--$\mu$ parameter space characteristic of the
    $\mu > 0$ GGM model relevant to the photon+$j$ analysis.
    For both figures, the lower plot shows the ratio of observed data to the combined SM expectation,
    with the inner band representing the range of statistical uncertainty and
    the outer band (visible only in the highest \RT bin in the right-hand figure) the combined statistical 
    and systematic uncertainty.
    Events outside the range of the displayed region are included in the highest-valued bin. 
    }
    \label{fig:diphoton_meff}
\end{figure}

A summary of the selection requirements specific to each
of the diphoton SRs is presented in Table~\ref{tab:dp-signalregion},
to \HBNL, \HBNH, \HBPL and \HBPH in Table~\ref{tab:hb-signalregion},
and to the two photon+$\ell$ SRs in Table~\ref{tab:l-signalregion}.
After all selection requirements, the numbers
of events remaining in the various signal
regions are 0 (\BSL, \BSH) , 5 (\BWL), 1 (\BWH), 12 (\HBNL), 2 (\HBNH, \HBPL, \HBPH),
16 (\WNLSPe) and 10 (\WNLSPu).

\begin{table}[bp]
  \caption{Enumeration of the requirements
defining the four SRs developed for the diphoton
signature search.  }
  \begin{center}
  \begin{tabular*}{\textwidth}{@{\extracolsep{\fill}}lcccc} \hline\hline\noalign{\smallskip}
    {\bf Signal Region}     &   \BSL       &  \BSH        &      \BWL    &    \BWH    \\
      \noalign{\smallskip}\hline\noalign{\smallskip}
    Number of photons ($\ET$ [GeV]) & $>1$ ($>75$)& $>1$ ($>75$)& $>1$ ($>75$)& $>1$ ($>75$) \\
    $\met $ [GeV]           &  $> 150$    & $> 250$     & $> 150$     & $> 200$    \\
    $\HT$  [GeV]             &    ...       &   ...        & $> 600$     & $> 400$   \\
    $\MEFF$ [GeV]            &  $> 1800$ & $> 1500$  &  ...     & ...   \\
    $\dphijm$ (Number of leading jets) & $> 0.5$ (2) & $> 0.5$ (2) & $> 0.5$ (2) & $> 0.5$ (2) \\
    $\dphim $               &  ...         & $> 0.5$     &   ...        &  $> 0.5$   \\
      \noalign{\smallskip}\hline\hline\noalign{\smallskip}
  \end{tabular*}
  \label{tab:dp-signalregion}
  \end{center}
\end{table}

\begin{table}[bp]
  \caption{Enumeration of the requirements
defining the four SRs developed for the photon+$b$ and photon+$j$
signature searches. 
}
  \begin{center}
  \begin{tabular*}{\textwidth}{@{\extracolsep{\fill}}lcccc} \hline\hline\noalign{\smallskip}
    {\bf Signal Region}       &    \HBNL       &     \HBNH      & \HBPL          & \HBPH           \\
      \noalign{\smallskip}\hline\noalign{\smallskip}
    Number of photons   ($\ET$ [GeV]) &$>0$ ($>125$)  &$>0$ ($>150$)  & $1$ ($>125$) & $1$ ($>300$)  \\
    $\met $ [GeV]           & $> 100$       & $> 200$       & $> 200$       & $> 300$        \\
    $\HT$ [GeV]          &   ...          & $> 1000$      & ...            & $> 800$        \\
    Number of jets (number of $b$-jets)   & $2-4$ ($> 1$) & $> 3$ ($> 0$) & $> 3^a$ & $> 1^a$ \\
    Number of leptons             &  0            &   ...          & 0             & 0              \\
    $M_{bb}$  [GeV] & $75-150$      & ...            & ...            & ...             \\
    \MTg [GeV]            & $> 90$        &  $> 90$       & ... & ...       \\
    $\dphijm$ (number of leading jets) & $> 0.3$ (2) & $> 0.3$ (4) & $> 0.4$ (2) & $> 0.4$ (2)   \\
    $\RT$                   &    ...         &   ...          & $< 0.85$      &      \\
    $\dphijg$               &    ...         &   ...          & ...            &  $< 2.0$ \\
      \noalign{\smallskip}\hline\hline\noalign{\smallskip}
  \end{tabular*}
  \label{tab:hb-signalregion}
  \end{center}
\vspace{-10pt}
{\small $^a$ For \HBPL and \HBPH, the two leading jets are required to have $\pt > 100$ and $\pt > 40$ GeV, respectively.}
\end{table}

\begin{table}[bp]
  \caption{Enumeration of the requirements
defining the two SRs developed for the photon+$\ell$
signature search.
}
  \begin{center}
  \begin{tabular*}{\textwidth}{@{\extracolsep{\fill}}lcc} \hline\hline\noalign{\smallskip}
    {\bf Signal Region}      &    \WNLSPe       &     \WNLSPu        \\
      \noalign{\smallskip}\hline\noalign{\smallskip}
    Number of photons ($\ET$ [GeV]) & $>0$ ($>125$)  & $>0$ ($>125$)    \\
    $\met $ [GeV]           &  $> 120$    & $> 120$         \\
    \HTj [GeV]      & $< 100$   & $< 100$          \\
    Number of leptons              & $> 0$ (e) & $> 0$ ($\mu$)       \\
    $|M_{e\gamma}-M_Z|$ [GeV]  & ($ > 15$)      &  ...         \\
    \MTl [GeV]         & $> 120$        &  $> 120$           \\
    $\Delta R(\ell,\gamma)$     &  $> 0.7$     & $> 0.7$   \\
      \noalign{\smallskip}\hline\hline\noalign{\smallskip}
  \end{tabular*}
  \label{tab:l-signalregion}
  \end{center}
\end{table}

\newpage
\section{Background estimation}
\label{sec:backgrounds}

Backgrounds to the various SRs arise from a number of sources, including processes
such as radiative vector boson production that generate real photons in combination with energetic neutrinos,
as well as events in which one or more energetic jets or electrons are misidentified as a photon.
While these sources contribute generically to all four signatures explored in
this study, the differing definitions of each of the associated SRs lead to, in many cases, significant
differences in the manner in which the contributions of these various background sources are estimated.
In the following, the methodology of the background estimation for each of the four experimental
signatures is discussed, and the resulting background estimates, broken down by source, are tabulated.
For the estimation of background contributions that rely upon MC simulation, either directly
or through the estimation of ``transfer factors'' relating the background content of control regions to
that of corresponding SRs, the effect of MC modeling uncertainties have been considered; in general,
these uncertainties are found not to be dominant contributions to the overall uncertainty in the background
estimates. Background models are confirmed in validation regions (VRs) with selection criteria closely 
related to those of the corresponding SR, but with one or more selection criteria modified to suppress
the contribution of possible GGM signal to the VR.

\subsection{Backgrounds to the diphoton analysis}
\label{sec:sub_binoback}

Backgrounds from SM contributions to the four diphoton SRs 
are grouped
into three primary components.  The first of these, referred
to as ``QCD background,'' arises from a mixture of
processes that include $\gamma \gamma$ production as well as 
$\gamma$ + jet and multijet events with at least one jet misreconstructed as a
photon.  The second background component, referred to as ``EW background,'' is due to $W+X$
(here ``$X$'' can be any number of jets, accompanied
by no more than one photon; the two-photon case is treated separately)
and \ttbar events, with a smaller contribution arising from $Z+X$ events.
These events tend to include final-state neutrinos that produce
significant \MET. In both cases, EW background events entering the signal regions
generally have at least one electron misreconstructed as a photon.
The QCD and EW backgrounds
are estimated through the use of dedicated control samples of data events.

The third background component,
referred to as ``irreducible,''
consists of $W$ and $Z$ bosons produced in association with two
real photons, with a subsequent decay into one or more neutrinos.
For this background, the $\Wgg$ component dominates, and
requires corrections to its LO contribution that
are both large and rapidly varying across the phase space of
the $\Wgg$ (plus possible additional jets) process~\cite{Bozzi:2011wwa}. Thus
a data-driven approach was developed to constrain the
$\Wgg$ contribution to the four SRs. The \Zgg contribution is estimated directly from the MC simulation.

The QCD background to \BSL, \BSH, \BWL and \BWH 
is expected to arise from events with two real, isolated photons (diphoton QCD events)
unaccompanied by any additional electroweak bosons, 
and from events with a single real, isolated photon and a jet whose
fragmentation fluctuates in such a manner as to cause it to be misidentified
as a second isolated photon (``photon+jet'' QCD events). A contribution from 
dijet QCD events is found to be small and largely incorporated into the
photon+jet background estimate.
To estimate the photon+jet contribution
a ``\QCDg control sample'' is identified
within the diphoton-trigger data sample by selecting events for which one photon candidate satisfies
the tight selection criterion, while the other
satisfies the loose but not the tight photon
criterion. 
\QCDg control sample events containing electrons are vetoed to reduce contamination from
$W\to e\nu$ decays.
Studies with MC simulated samples as well as \MET and \HT sideband data
suggest that the \MET distribution of this control sample adequately
reproduces the \MET distribution of the QCD background in the high-\MET
region used for the signal selection.
A diphoton MC sample is used for the estimation of the diphoton contribution to the QCD background.

The \HT, \MEFF, \dphijm and \dphim requirements associated with each of the four SRs are applied
to the \QCDg control and diphoton MC samples, and the resulting samples
are scaled so that the combination of the two samples exactly reproduces the number of
observed diphoton events (for the given SR) in the region $0 < \MET < 60$ GeV, and with the
diphoton MC sample providing a specified fraction of the total event count in this region. 
As suggested by the independent ATLAS $H \rightarrow \gamma\gamma$~\cite{Aad:2014lwa} and
isolated photon pair cross-section~\cite{Aad:2012tba} analyses, 
this fraction is set to 75\%, although in this analysis
a range
between 50\% and 100\% is adopted to reflect the degree of uncertainty in this fraction.
The resulting QCD-background estimate, for each of the four binolike SRs, is then
obtained by summing the scaled number of combined \QCDg control and diphoton MC events with \MET
above the minimum requirement for the given SR. Additional sources of systematic
uncertainty on the QCD-background estimate include its dependence on the low-\MET
region used to scale the diphoton MC and \QCDg control samples, and the effect of
possible mismodeling of the \dphijm and \dphim distributions of the QCD background 
by the \QCDg control sample. Including both systematic uncertainty and the
statistical uncertainty associated with the limited number of events in the \QCDg control
and diphoton MC samples, the result for the QCD background and its overall uncertainty
is shown in Table~\ref{tab:diphoton_back}.

The QCD-background model is validated by comparing the observed numbers
of events to the total expected SM background in bins of \unit[300]{GeV}
in \HT for the sideband region $100 < \MET < 150$ GeV, for which event
rates are expected to be dominated by the QCD background. The observed event
rate tends to be somewhat lower than that predicted by the overall background
model, although it is within 1 standard deviation of the overall background model
uncertainty for all \HT bins.

The EW background, arising predominantly from $W+X$ and \ttbar
events, is estimated via an ``electron-photon'' control sample
composed of events with at least one isolated tight photon and one isolated electron, each with
$\ET > 75$ GeV, and scaled by the probability for such an electron to
be misreconstructed as a tight photon, as estimated from a ``tag-and-probe'' study of the
$Z$ boson in the $ee$ and $e\gamma$ sample. The electron-to-photon scale factor varies
between 1.9\% ($0 < |\eta| < 0.6$) and 3.7\% ($1.52 < |\eta| < 1.81$),
since it depends on the amount of material in front of the calorimeter.
Events with two or more tight photons are vetoed from the control sample to preserve
its orthogonality to the signal sample. In the case of more than one electron,
the one with the highest \pt is used.
Including systematic uncertainties of $\pm 25$\% each, associated with a possible
\pt dependence of the scale factor and a possible overlap between the
QCD and EW background estimates, leads to the estimates
for the EW background to the four diphoton SRs shown in Table~\ref{tab:diphoton_back}.

The irreducible background is composed of two distinct components: diphoton production 
in association with either a $W$ or $Z$ boson. The latter contribution is relatively 
small and is sufficiently well understood to allow the use of the 
MC simulation, with a total cross section scaled to that of Ref.~\cite{Bozzi:2011en}, to
directly estimate the \Zgg contribution to the four SRs. The value of this
estimate is shown is Table~\ref{tab:diphoton_back}; the uncertainty is dominated
by a $\pm 50$\% uncertainty on the \Zgg cross section of Ref.~\cite{Bozzi:2011en}
that arises from the variation of the factorization and renormalization scales
used to quantify the uncertainty due to missing higher-order processes.

The \Wgg background to the four SRs is estimated using
a lepton-diphoton (\lgg) CR. To enhance
the contribution of \Wgg and ensure that the \lgg CR
is exclusive of the four SRs, the photon \pt requirement is lowered to
50 GeV and a requirement of $50 < \MET < 150$ GeV is imposed. To ensure that
the CR sample arises from the same region of the \Wgg process phase space
as the expected background, a further requirement that the transverse momentum 
of the \lgg system be greater than 100 GeV is imposed. 
A total of seven events is observed in the CR, for which MC simulation suggests that 2.2
are expected to arise from SM sources other than \Wgg. When setting limits
on contributions from new physics in the four SRs, a simultaneous fit
to the CR and the signal region under study is performed,
allowing both the signal and \Wgg contributions to float to their best-fit values.
When setting model-dependent limits, the fit also takes into account a possible signal contribution to the \lgg CR,
which can be significant for the electroweak-production models in the
case that the \neutralino mass is light. 
In the limit that no GGM signal contributes to the \lgg control region,
an enhancement factor of 2.3 must be applied to the \Wgg MC sample to achieve
agreement between the MC simulation and data in the \lgg control region.
The resulting \Wgg-background estimate in each of the four SRs, under
the assumption that there is no signal contribution to the \lgg CR, 
is shown in Table~\ref{tab:diphoton_back}; the uncertainty
is dominated by that of the limited number of events in the \lgg CR. 
Also shown is the combined background estimate, including uncertainty,
from all four sources.

\begin{table}[bp]
    \caption{The expected and observed numbers of events for
      the four diphoton signal regions. The quoted errors are the combined
      statistical and systematic uncertainties.}
  \begin{center}
  \begin{tabular*}{\textwidth}{@{\extracolsep{\fill}}lcccc} \hline\hline\noalign{\smallskip}
      {\bf Signal Regions} & \BSL & \BSH & \BWL & \BWH  \\
      \noalign{\smallskip}\hline\noalign{\smallskip}
      Expected background events & $0.06^{+0.24}_{-0.03}$   & $0.06^{+0.24}_{-0.04}$ & $2.04^{+0.82}_{-0.75}$ & $1.01^{+0.48}_{-0.42}$ \\
        QCD & $0.00^{+0.24}_{-0.00}$ & $0.00^{+0.24}_{-0.00}$ & $0.32^{+0.45}_{-0.32}$& $0.22^{+0.33}_{-0.22}$ \\
        EW  & $0.02\pm0.02$  & $ 0.0 \pm 0.0$           & $0.64\pm0.27$  & $0.13\pm0.08$ \\
 $(W\to\lnu)\gamma\gamma$ & $0.04\pm0.02$  & $0.05\pm0.04$ & $1.01\pm0.62$ & $0.53\pm0.34$ \\
$(Z\to\nu\nu)\gamma\gamma$  & $0.00\pm0.00$  & $0.01\pm0.01$  & $0.07\pm0.04$  & $0.13\pm0.07$ \\
     Observed events & 0 & 0 & 5 & 1 \\
 \hline\hline
   \end{tabular*}
   \label{tab:diphoton_back}
  \end{center}
\end{table}

\subsection{Backgrounds to the photon+$b$ analysis}
\label{sec:sub_higgs_bino_m_back}

For both \HBNL and \HBNH, which include a requirement of
at least one $b$-jet, backgrounds arise from two predominant
sources: from leptonic decays of real or virtual $W$ bosons accompanied by the
production of $b$-quark pairs, including those arising in \ttbar
events [``\wlnu'' backgrounds]; and from events containing no electroweak bosons
or top quarks (QCD backgrounds). \wlnu background events are further classified
according to the origin of the high-energy isolated photon. Contributions from \wlnu backgrounds for
which the photon arises from the misidentification of an electron
are estimated via a control sample for which the photon requirement is replaced by an
electron requirement, scaled by an electron-to-photon
misidentification probability; this approach is similar in nature to that of the diphoton analysis.
Estimates of this component of the background to \HBNL and \HBNH are
shown in Table~\ref{tab:background_photon+b}; the quoted uncertainty arises from
the limited number of events in the control sample, as well as systematic uncertainty associated
with the possible \pt dependence of the electron-to-photon misidentification-rate scale factor.

\begin{table}[bp]
    \caption{The expected and observed numbers of events for
      the two photon+$b$ signal regions. The quoted errors are the combined
      statistical and systematic uncertainties. }
  \begin{center}
  \begin{tabular*}{\textwidth}{@{\extracolsep{\fill}}lcc} \hline\hline\noalign{\smallskip}
    {\bf Signal Regions} & \HBNL & \HBNH   \\
      \noalign{\smallskip}\hline\noalign{\smallskip}
      Expected background events & $18.8 \pm 5.3 $ &     $3.82 \pm     1.25$  \\ 
$e \rightarrow\gamma$  &     $3.2 \pm     0.4$  &     $0.18 \pm     0.08$ \\
\wlnu  &     $12.6 \pm      4.9$   &     $3.35 \pm     1.05$ \\
QCD  &      $2.3 \pm      2.1$  &    $0.00 \pm     0.65$ \\
$Z \rightarrow\nu\nu$ &     $0.8 \pm     0.4$  &     $0.29 \pm     0.15$  \\
Observed events & 12 & 2 \\
 \hline\hline
   \end{tabular*}
   \label{tab:background_photon+b}
  \end{center}
\end{table}

Contributions from \wlnu backgrounds for which the photon is real,
or for which the photon arises from a misidentified jet or $\tau$ lepton,
are estimated via lepton-enriched CRs that constrain the normalization
of MC samples used to simulate contributions from these two sources.
Separate control regions CR$^{\mathrm {lep}}_{\rm L}$ and  CR$^{\mathrm {lep}}_{\rm H}$ are defined for the low- and
high-neutralino-mass SRs by requiring a lepton in addition to the
requirements already imposed to define the SRs. In addition, in order to
increase the number of events in the CR, the
\MET requirement is reduced, the $M_{bb}$ requirement is removed (for the \HBNL analysis),
and the \HT requirement is relaxed (for the \HBNH analysis). 
Events in these two CRs are expected to be dominated by \ttbar, $\ttbar \gamma$
and $W \gamma$ production, as is expected for the corresponding 
background contributions to the SRs, and
any overlapping phase space is subtracted as part of the background estimation.
Including all SM sources, a total of 14.5 (58.0) events are
expected in the   CR$^{\mathrm {lep}}_{\rm L}$ ( CR$^{\mathrm {lep}}_{\rm H}$) control regions, to be compared to
an observation of 18 (61) events. Scaling the combined SM MC samples by
these ratios of data to expectation,
after having subtracted the contributions estimated by other techniques,
yields the SR background estimates
shown in Table~\ref{tab:background_photon+b}.
It is found that the data-to-expectation scale factor is somewhat
dependent upon the requirements used to define the lepton-enriched 
CRs; these variations are included in the systematic
error on the resulting SR background prediction.

The QCD background is estimated via the definition of a two-dimensional
signal- and control-sample grid (the ``ABCD'' method). For the \HBNL analysis, three
control samples are defined by requiring only a single tagged
$b$-jet, by requiring that $\MET < 75$ GeV, or by requiring both
of these SR modifications. 
For the \HBNH analysis, three similar control samples are defined
by requiring that no jet be identified as a $b$-jet, 
by requiring that $\MET < 150$ GeV, or by requiring both
of these SR modifications. A transfer factor is calculated
by taking the ratio of the number of events with only the \MET requirement changed
to the number of events with both the \MET and $b$-jet requirements changed.
Assuming that the 
relaxation of the $b$-jet requirement is uncorrelated with the
relaxation of the \MET requirement, 
the number of QCD-background events
in the SR can then be estimated by scaling, by this transfer factor, the number of events
with only the $b$-tag requirement changed.
This scaling is done only after subtracting the number
of events expected to come from sources other than those that produce QCD-background events
from each of the control samples.
To avoid the biasing effects of possible correlations 
between the relaxation of the $b$-jet requirement and 
the \MET requirement, for the \HBNL (\HBNH) analysis 
events are binned in \MET
and weighted bin by bin in the ratio of the number of events in the 2-tag
(1-tag) region to the number of events in the 1-tag (0-tag) region
in the $\gamma$+jet MC sample. 
The resulting estimate of the small expected
QCD background in the two SRs is shown in 
Table~\ref{tab:background_photon+b}, with the systematic uncertainty
dominated by the limited number of events to which the scale
factor is applied.

An additional background due to the production of a $Z$ boson 
that decays into two neutrinos, in 
association with a photon and a $b$-jet, is estimated
directly from the MC simulation, and is tabulated in Table~\ref{tab:background_photon+b}.
For this final contribution, a 50\% scale error is assumed for the
overall rate of production for this process. The combined background 
from all expected sources is also shown in Table~\ref{tab:background_photon+b}.

For both photon+$b$ SRs, the background model is validated in four
VRs, defined for \HBNL by requiring $75 < \MET < 100$ GeV,
by reversing the $M_{bb}$ requirement, by requiring $\MTg < 90$ GeV,
or by requiring $\dphijm < 0.3$, respectively. Since no $M_{bb}$
requirement is made for \HBNH, the second validation region is
instead defined by changing the \HT requirement to $500 < \HT < 1000$ GeV.
The observed numbers of events in the VRs are consistent with the predictions
of the overall background model. 

\subsection{Backgrounds to the photon+$j$ analysis}
\label{sec:sub_higgs_bino_p_back}

Backgrounds to the photon+$j$ analysis are expected to arise both from 
events with real photons as well as events for which an electron or a 
jet is misidentified as a photon. The former source is
expected to receive contributions from events for which a $W/Z$ boson, a single top quark,
or a \ttbar pair is produced in association with a real photon, with
neutrinos in the subsequent weak decays of these produced states 
providing significant \MET ($W\gamma$, $Z\gamma$ and $\ttbar\gamma$ background).
Events with real photons can also contribute to the background to the photon+$j$ analysis when significant \MET 
arises from instrumental sources (QCD background).
The $W\gamma$, $\ttbar\gamma$ and QCD backgrounds are estimated by scaling a
corresponding MC sample to match the observed event count in a corresponding CR
enriched in the given background process but otherwise kinematically 
similar to the corresponding SR. The MC simulation is then used
to provide an estimate of the expected background in the \HBPL and 
\HBPH SRs. Smaller contributions from single-top$+\gamma$ and $Z$$\gamma$ are estimated 
directly from the MC simulation.

The QCD-background CR
is defined by changing the \HBPL and \HBPH \MET requirements to instead select events with
$\MET < 50$ GeV, but leaving all other selection requirements unchanged, 
providing a region dominated by real photons arising from radiative
QCD processes. 
The $W\gamma$-background CR
is defined by requiring, in addition to the other \HBPL and \HBPH requirements, that there be a single 
identified isolated lepton (electron or muon) and no $b$-jet in the event. 
The $\ttbar\gamma$-background CR
is defined similarly, but requires instead at least one $b$-jet. In both cases, 
in order to increase the number of events in the CR the \MET requirement is changed 
to $100 < \MET < 200$ GeV. The event counts in the resulting QCD, $W\gamma$ and $\ttbar\gamma$ CRs are
used to scale the $\gamma$+jet, $W\gamma$ and $\ttbar\gamma$ MC samples, respectively, after applying a selection identical to
that of the corresponding CR. 
The scale factors are determined in a simultaneous fit to all CRs,
taking into account mutual cross contamination between the different backgrounds.
Estimates for the contributions of
all three of the real-photon backgrounds are shown in Table~\ref{tab:background_photon+j}.
Systematic uncertainty on the scale factor
is dominated by the theoretical uncertainties on the relevant MC samples, related in turn to the PDF choice 
and the renormalization and factorization scales. 

As in the other analyses, backgrounds from events
for which electrons are misidentified as photons are estimated
by identifying a control sample of events through the application of a set of selection requirements
that are identical to those of the given SR, but with
a requirement that the event have an electron that replaces the required photon.
The estimate of the background in the 
SR (\HBPL or \HBPH) is then, as in the other analyses,
derived by scaling each event in the control sample by an
$\eta$-dependent electron-to-photon misidentification factor.
The resulting background estimates are displayed in Table~\ref{tab:background_photon+j}.

Finally, the contribution of a background due to events for which the selected
photon arises from the misidentification of a jet is
estimated by determining the jet-to-photon misidentification rate 
from the observed isolation-energy distribution of energy in a 
cone of size $\Delta R = 0.2$ surrounding the
energy deposition in the calorimeter associated with the photon. 
The isolation-energy distribution for real photons is modeled with electrons
from $Z$ boson decays, while that of misidentified jets
is modeled with a sample of events for which there is a ``pseudophoton.'' A
pseudophoton is defined to be an object that
passed all loose photon selection requirements, as well as all tight
photon selection requirements except one or more from a set of
four that relate to the shape of the deposition in the finely granulated
front portion of the EM calorimeter. 
The fraction of misidentified jets within the tight, isolated photon sample is determined with
a control sample composed of events with tight, isolated photons with
$\pt > 125$ GeV,
as well as a relaxed \MET requirement of $50 < \MET < 150$ GeV
and an intermediate requirement of $\HT > 600$ GeV. 
The photon isolation-energy distribution
of this control sample is fit to establish the relative amounts
of these two sources (real photons and misidentified jets), with the
misidentification fraction taken to be the relative integrals
of the isolation-energy distributions of the misidentified and total contributions in the region
for which the isolation energy is less than 5 GeV.
The estimation of the jet-misidentification background in each signal region and control sample (as well
as for the validation regions described below) is then obtained
by scaling the observed number of events in each region or sample by the jet-misidentification factor. 
The number of misidentified jets 
is then parametrized as a function of \MET by fitting the \MET dependence of the estimated misidentified-jet contribution
in the range $\MET<200$ GeV. The estimates 
in the \HBPL and \HBPH signal regions are then extracted by integrating the fit function 
over the relevant \MET range. The result for the contribution of 
the jet-misidentification backgrounds for each SR is shown in Table~\ref{tab:background_photon+j}.
Systematic uncertainties arise due to the uncertainties in the combined fit used to derive the 
misidentification factor, for which the parameters
of the signal and background templates are allowed to vary within their uncertainties, and from the uncertainties of the extrapolation 
fit used for the estimation of the SR contamination.

The background model is validated by comparing expected and observed event rates in
several VRs. For \HBPL, this includes three VRs for which the $\dphijm$ is reversed, 
\MET is required to be within an intermediate range of $75 < \MET < 150$ GeV,
and for which the \RT requirement is reversed. For \HBPH two VRs are made use of, including
one for which $\dphijm$ is reversed and another that requires that $400 < \HT < 800$ GeV.
Good agreement is observed between the number of expected and observed events in all
five VRs.

\begin{table}[bp]
    \caption{The expected and observed numbers of events for
      the two photon+$j$ signal regions. The quoted errors are the combined
      statistical and systematic uncertainties.}
  \begin{center}
  \begin{tabular*}{\textwidth}{@{\extracolsep{\fill}}lcc} \hline\hline\noalign{\smallskip} 
      {\bf Signal Regions} & \HBPL & \HBPH  \\
      \noalign{\smallskip}\hline\noalign{\smallskip}  
      Expected background events         & $1.27 \pm 0.43$              &  $0.84 \pm 0.38$              \\
        $W + \gamma$          & $0.13 \pm 0.12$              &  $0.54 \pm 0.28$              \\
        $Z + \gamma$          & $0.03_{-0.03}^{+0.05}$              &  $0.21_{-0.21}^{+0.23}$              \\
        $t\bar{t}$ + $\gamma$          & $0.64 \pm 0.40$              &  $0.05 \pm 0.05$              \\
        Single-$t$ + $\gamma$          & $0.06 \pm 0.02$              &  $0.03 \pm 0.01$              \\
        $\gamma$ + jet  (QCD background)    & $0.00_{-0.00}^{+0.06}$              &  $0.00 \pm 0.00$              \\
        $e\rightarrow\gamma$               & $0.38 \pm 0.10$              &  $0.00 \pm 0.00$              \\
        $j\rightarrow\gamma$                & $0.02_{-0.02}^{+0.08}$              &  $0.00_{-0.00}^{+0.08}$        \\      
Observed events & 2 & 2 \\
 \hline\hline
   \end{tabular*}
   \label{tab:background_photon+j}
  \end{center}
\end{table}

\subsection{Backgrounds to the photon+$\ell$ analysis}
\label{sec:sub_wino_nlsp_back}

Backgrounds to the photon+$\ell$ analysis (\WNLSPe and \WNLSPu) 
are expected to arise primarily from events with
hard photons produced in association with
electroweak bosons ($W\gamma$ or $Z\gamma$) and top quarks ($\ttbar\gamma$),
and events containing $W$ bosons 
or semileptonically decaying top quarks 
for which an accompanying jet is
misidentified as a photon (jet-to-photon events). 
Lesser contributions are expected to arise
from \ttbar events and events containing two electroweak bosons that produce
two final-state leptons, one of which is an electron that is subsequently
misidentified as a photon.
As in the other analyses, data-driven techniques making use of CRs similar to
but exclusive of the SRs, or control samples appropriate for assessing
jet-to-photon and electron-to-photon misidentification rates, are used to estimate or constrain the
primary backgrounds, while lesser backgrounds are estimated
directly from MC simulation.

The most prevalent background in the photon+$\ell$ sample is expected
to arise from $W\gamma$ events. A $W\gamma$ CR is defined by requiring an
isolated electron or muon, and by requiring in addition that
$45 < \MET < 100$ GeV and 
$35 < \MTl < 90$ GeV, but otherwise requiring that the sample
satisfy the \WNLSPe and \WNLSPu criteria. Transfer factors relating
the number of events observed in the $W\gamma$ CR 
to the number of  $W\gamma$ events expected in the SRs are estimated,
separately for the electron and muon contributions, from the $W\gamma$ MC simulation.
Systematic uncertainties on the resulting $W\gamma$-background estimate for
the two SRs arise from the scale and PDF uncertainties associated with the transfer factors.
A somewhat lesser contribution from $\ttbar\gamma$ events is estimated directly
from the MC simulation, with uncertainties arising from imprecise knowledge of the strong-interaction
scale and the rate of final-state photon production into the acceptance of the SRs.
A smaller background contribution from $Z\gamma$ events is estimated directly
from the MC simulation, with an uncertainty of $\pm 50$\% assumed
for the production rate into the region of the $Z\gamma$ phase space that 
populates the photon+$\ell$ SRs.

As in the other analyses, a potentially sizable contribution to the photon+$\ell$ sample arises from
jet-to-photon misidentification. 
The contribution of SR events
arising from jet-to-photon misidentification is estimated by
exploring the isolation-energy distribution of events in an extended
$W\gamma$ control sample for which the requirement on isolation energy has been removed.
Isolation-energy distribution templates for true photons and for jets 
misidentified as photons are developed in the manner described for the photon+$j$ analysis.
A fit is then performed on the isolation-energy distribution 
of the extended $W\gamma$ control sample
to estimate the number of events in the isolated (isolation energy
less than \unit[5]{GeV}) $W\gamma$ CR that arise from jets
misidentified as photons. A scale factor is
defined as the ratio of the estimated number of events in the isolated $W\gamma$ CR
arising from misidentified jets to that expected from the $W$+jets and
semileptonic \ttbar MC simulations. A data-driven
estimate of the number of events arising from misidentified jets in
\WNLSPe and \WNLSPu is then derived by multiplying the number of
such events expected from the combination of the $W+$jets and 
semileptonic \ttbar MC simulations by this scale factor.
Because the MC simulation is relied upon to propagate the background estimate from
the control sample into the SR, uncertainties on the jet-to-photon misidentification background
arise due to imprecise knowledge of the proton PDFs and strong-interaction scale.
An additional uncertainty is assigned based on the difference between the scale factors
determined for the separate electron and muon samples. 

A final significant source of background is expected to arise from 
\ttbar events, single-top events, and events containing two electroweak bosons that produce
two final-state leptons, one of which is an electron that is subsequently
misidentified as a photon. The contribution from these backgrounds is
estimated from MC simulation, applying a correction based on the relative electron-to-photon misidentification 
rate between data and MC simulation. This correction is 
determined from $Z \rightarrow e^+e^-$ events as described above for other analyses.
In addition to the
uncertainty in the measurement of the misidentification rate,
uncertainties in the estimate arise from PDF and scale uncertainty
in the \ttbar production process as well as an assumption
of a $\pm 50$\% uncertainty in the rate of single-top and diboson production.

All other sources of background, including those from
$Z$+jet, $\gamma$+jet
and $\gamma\gamma$ production, are expected to contribute only
minimally to the total SR backgrounds. In particular, a potential
background from $\gamma$+jet events arising from jet-to-lepton misidentification
is estimated using a matrix method (as described in Ref.~\cite{ref:matrix_method}) making use
of a control region incorporating nonisolated lepton candidates, and is
found to contribute 0.1 events to 
the overall background estimate for each of the \WNLSPe and \WNLSPu
samples.
A summary of the resulting background estimates for the \WNLSPe and \WNLSPu
SRs is shown in Table~\ref{tab:background_photon+l}, broken down
by source. 

The background model is validated for each SR by comparing expected and observed event rates in
two VRs. An $M_\mathrm{T}$ VR is defined by relaxing the \MTl requirement to $35 < \MTl < 90$ GeV;
to increase the number of events in this VR, the \MET requirement is also relaxed to $\MET > 100$ GeV.
A \MET{}  VR is defined by relaxing the \MET requirement
to $45 < \MET < 100$ GeV while leaving the \MTl requirement unchanged. For
the electron and muon channels combined, the number of events in the \met{} VR
is observed to be somewhat less than that expected for the background model,
although still within 2 standard deviations of the combined statistical
and systematic uncertainty. Good agreement is found for the $M_\mathrm{T}$ VR.

\begin{table}[bp]
    \caption{The expected and observed numbers of events for
      the two photon+$\ell$ signal regions. The quoted errors are the combined
      statistical and systematic uncertainties. 
      The contribution from
      the $Z\gamma$ process arises from events for which one of the 
      leptons from the $Z \rightarrow \ell^+ \ell^-$ decay is either
      missed or badly mismeasured. The likelihood of this occurring
      is significantly greater for muons than electrons.}
  \begin{center}
  \begin{tabular*}{\textwidth}{@{\extracolsep{\fill}}lcc} \hline\hline\noalign{\smallskip} 
    {\bf Signal Regions} & \WNLSPe & \WNLSPu   \\
      \noalign{\smallskip}\hline\noalign{\smallskip}
      Expected background events   & $10.5 \pm 1.4$  & $14.1 \pm 1.5$  \\
   $W \gamma$               & $6.7 \pm 1.2$   & $8.8 \pm 1.3$   \\
   $\ttbar \gamma$             & $1.4 \pm 0.6$   & $1.7 \pm 0.7$   \\
   $Z \gamma$               & $0.0 \pm 0.0$   & $1.2 \pm 0.6$   \\
   Jet $\rightarrow \gamma$ & $1.5 \pm 1.0$   & $1.2 \pm 0.7$   \\
   $e\rightarrow \gamma$    & $0.7 \pm 0.2$   & $0.8 \pm 0.3$   \\
   Other sources            & $0.3 \pm 0.1$   & $0.4 \pm 0.2$   \\
Observed events & 16 & 10 \\
 \hline\hline
   \end{tabular*}
   \label{tab:background_photon+l}
  \end{center}
\end{table}

\section{Signal efficiency and systematic uncertainty}
\label{sec:sig_eff}

GGM signal acceptances and efficiencies are estimated using MC simulation for each simulated
point in the gluino-bino, wino-bino, higgsino-bino, and wino-NLSP parameter spaces,
and vary widely across the regions of these spaces relevant to establishing the limit contours 
presented below. The product of acceptance times efficiency tends to be greatest (10\%--25\%)
when the masses of 
both the produced and the NLSP states are largest, leading to large amounts of both visible energy and
missing transverse momentum that would clearly distinguish signal from background events. However, 
for the lower accessible mass scales associated with electroweak production, and particularly for
the case of a low-mass NLSP, the product of
acceptance times efficiency can be significantly smaller. 
For example, for the region relevant to establishing limits at low values of
$\mu$, the efficiency of the \HBPL analysis is less than 0.1\%,
leading to a relatively modest lower limit
on the mass of produced SUSY states.

Making use of a bootstrap method~\cite{ATLAS-CONF-2012-048},
the efficiencies of both the single photon and diphoton triggers are determined
to be greater than 99\%, with an uncertainty of less than 1\%. 

The reconstruction efficiency for tight, isolated photons is estimated with
complementary data-driven methods~\cite{ATLAS-CONF-2012-123}. 
Photons identified kinematically as having come from radiative decays of a $Z$ boson
($Z \rightarrow \ell^+ \ell^- \gamma$ events)
are used to study the photon reconstruction efficiency as a function of $\pt$ and $\eta$. 
Independent measurements making use of a tag-and-probe approach with 
$Z \to ee$ events, with one of the electrons
used to probe the calorimeter response to electromagnetic depositions, also provide 
information about the photon reconstruction efficiency. For photons with $\ET > 75$ GeV,
the identification efficiency in the range $0 < |\eta| < 1.81$ is greater than 95\%;
for the range $1.81 < |\eta| < 2.37$ the efficiency is approximately 90\%. The 
uncertainty in the efficiency 
also varies with $|\eta|$, and lies between $\pm (1$--$2)\%$

The isolated electron efficiency is also estimated
using tag-and-probe methods, making use of samples of $Z \to ee$
and $J/\psi \to ee$ events as described in
Refs.~\cite{Aad:2014fxa,ATLAS-CONF-2014-032}. The efficiency and its
uncertainty are estimated
as a function of electron \pt and $\eta$, leading to an overall uncertainty 
of $\pm 1.0$\% 
on the efficiency of the photon+$\ell$ analysis, the only analysis that explicitly
requires an electron. 
The muon identification uncertainty, estimated as described in Ref.~\cite{Aad:2014rra}, is found to contribute 
an uncertainty of only 0.4\% on the efficiency of the photon+$\ell$ analysis.

In portions of the GGM parameter space, uncertainties that vary
across the parameter space dominate the systematic uncertainty on
the signal acceptance times efficiency. These model-dependent uncertainties 
include those due to
uncertainties in the photon, electron and jet-energy scales, the $b$-jet tagging
efficiency, and the ``pileup'' uncertainty arising from the modeling of additional interactions
in the same or nearby bunch crossings. 

The electron and photon energy scale is
determined using samples of $Z \to ee$ and $J/\psi \to ee$ events~\cite{Aad:2011mk},
both of whose masses are known precisely and thus 
provide an accurate calibration signal for determination 
of the electromagnetic calorimeter response. Uncertainties arise
from imprecise knowledge of the material burden between the
IP and the face of the EM calorimeter.
The muon energy scale and uncertainty are similarly estimated
with calibration samples of $Z \to \mu\mu$, $\Upsilon \to \mu\mu$ 
and $J/\psi \to \mu\mu$ events~\cite{Aad:2014rra}.

The jet-energy scale is established via the 
propagation of single-particle test-beam measurements of the
calorimeter response through simulations of jets arising from
$pp$ collisions~\cite{Aad:2011he,Aad:2012vm}. The jet-energy scale 
uncertainty is constrained by
the study of momentum imbalance in dijet events~\cite{Aad:2012ag}, as well as
from an assessment of the effect of uncertainties in the modeling of jet properties with
MC simulations, and from uncertainties in the modeling of the varying response to differing
jet flavor composition.

Uncertainties in the values of whole-event observables, such as \MET and \HT, arise from 
uncertainties on the energy of the underlying objects from which they are constructed. In addition, 
the \MET observable receives a contribution from calorimetric energy deposits not associated with 
any of the reconstructed objects in the event. Uncertainties on the energy scale of these unassigned 
contributions are found to contribute negligibly to the overall uncertainty on the value of 
the \MET observable.

The uncertainty due to pileup is estimated by varying the distribution of the number
of interactions per bunch crossing overlaid in the simulation by $\pm 10$\%.
The uncertainty on the $b$-tagging efficiency in the MC simulation 
is estimated from measurements of dedicated heavy-flavor calibration data samples.

In the regions of GGM parameter space relevant for establishing the exclusion
limits discussed below, and
excepting MC statistical uncertainty, the quadrature sum of the individual sources of systematic uncertainty on the signal reconstruction
efficiency for the diphoton, photon+$b$ and photon+$\ell$ analyses is
of order 10\%. For the photon+$j$ analysis the systematic uncertainty is somewhat 
larger---approximately 20\%--- due to an increased sensitivity to the jet-energy scale
and resolution associated with the multiple-jet requirement.

\section{Results}
\label{sec:results}

\begingroup
\renewcommand*{\arraystretch}{1.15}
\begin{table}[bp]
  \caption{Summary of the number of events expected from SM sources
($N_{\rm exp}^{\rm SM}$), and the observed number of events ($N_{\rm obs}$), for each of the ten SRs.
Also shown is the derived model-independent 95\% CL limit ($S_{\rm obs}^{95}$) on the number
of possible events from new physics, as well as both the observed ($\langle\epsilon{\rm \sigma}\rangle_{\rm obs}^{95}$) 
and expected ($\langle\epsilon{\rm \sigma}\rangle_{\rm exp}^{95}$) 95\% CL limit on the visible cross section from new physics.
Due to the discrete nature of the number-of-observed-events likelihood distribution in background-only
pseudoexperiments, when both the
expected number of background events and its uncertainty are close to zero
the expected limit is dominated by the case of zero observed events.
This leads to a very narrow
one-standard-deviation range for the expected limit for \BSL and \BSH.
   }
  \label{tab:mod_ind_lim}
\begin{center}
\setlength{\tabcolsep}{0.0pc}
\begin{tabular*}{\textwidth}{@{\extracolsep{\fill}}lccccc}
\noalign{\smallskip}\hline\hline\noalign{\smallskip}
{\bf Signal Region}                    & $N_{\rm obs}$ & $N_{\rm exp}^{\rm SM}$ & $S_{\rm obs}^{95}$ & $\langle\epsilon{\rm \sigma}\rangle_{\rm obs}^{95}$[fb] & $\langle\epsilon{\rm \sigma}\rangle_{\rm exp}^{95}$[fb]    \\
\noalign{\smallskip}\hline\noalign{\smallskip}
\BSL    & 0 &  $0.06^{+0.24}_{-0.03}$ & 3.0 & 0.15 & $0.15 \pm 0.01$ \\
\BSH    &  0  & $0.06^{+0.24}_{-0.04}$ & 3.0 & 0.15 &  $0.15 \pm 0.01$ \\
\BWL     &   5 & $2.04^{+0.82}_{-0.75}$ & 8.2 & 0.41 & $0.25^{+0.09}_{-0.06}$   \\
\BWH      &   1 &  $1.01^{+0.48}_{-0.42}$ & 3.7 & 0.18 & $0.18^{+0.07}_{-0.02}$   \\
\HBNL     & 12 & $18.8 \pm 5.4$ & 8.1 & 0.40 & $0.57^{+0.24}_{-0.16}$ \\
\HBNH   &  2 & $ 3.82 \pm 1.25$ & 4.0 & 0.20 & $0.27^{+0.09}_{-0.07}$ \\
\HBPL     &  2 & $1.27 \pm 0.43$ & 5.5 & 0.27 & $0.19^{+0.10}_{-0.06}$ \\
\HBPH  & 2 & $0.84 \pm 0.38$ & 5.6 & 0.28 &  $0.20^{+0.11}_{-0.05}$ \\
\WNLSPe & 16 & $10.5 \pm 1.4$ & 14.2 & 0.70 & $0.41^{+0.20}_{-0.12}$ \\
\WNLSPu & 10 & $14.1 \pm 1.5$ &  6.0 & 0.30 & $0.45^{+0.21}_{-0.14}$  \\
\noalign{\smallskip}\hline\hline\noalign{\smallskip}
\end{tabular*}
\end{center}
\end{table}
\endgroup

An accounting of events observed in each SR 
is shown in Table~\ref{tab:mod_ind_lim},
along with the size of the expected SM
background.
Comparisons of the \MET distribution between signal
and expected background is shown for several different SRs
in Figs.~\ref{fig:diphoton_met}--\ref{fig:photon_l_met}.
No evidence for physics beyond the SM is observed in any of the SRs.
The largest excess relative to the expected background is observed for the \BWL analysis;
considering both statistical and systematic uncertainty, and
assuming that all observed events are from SM sources,
an observation of five or more events over an expected background of
$2.04^{+0.82}_{-0.75}$ represents an upward fluctuation with
a probability of occurrence of approximately 6\%.

\begin{figure}[tp]
  \begin{center}
    \includegraphics[width=0.48\textwidth]{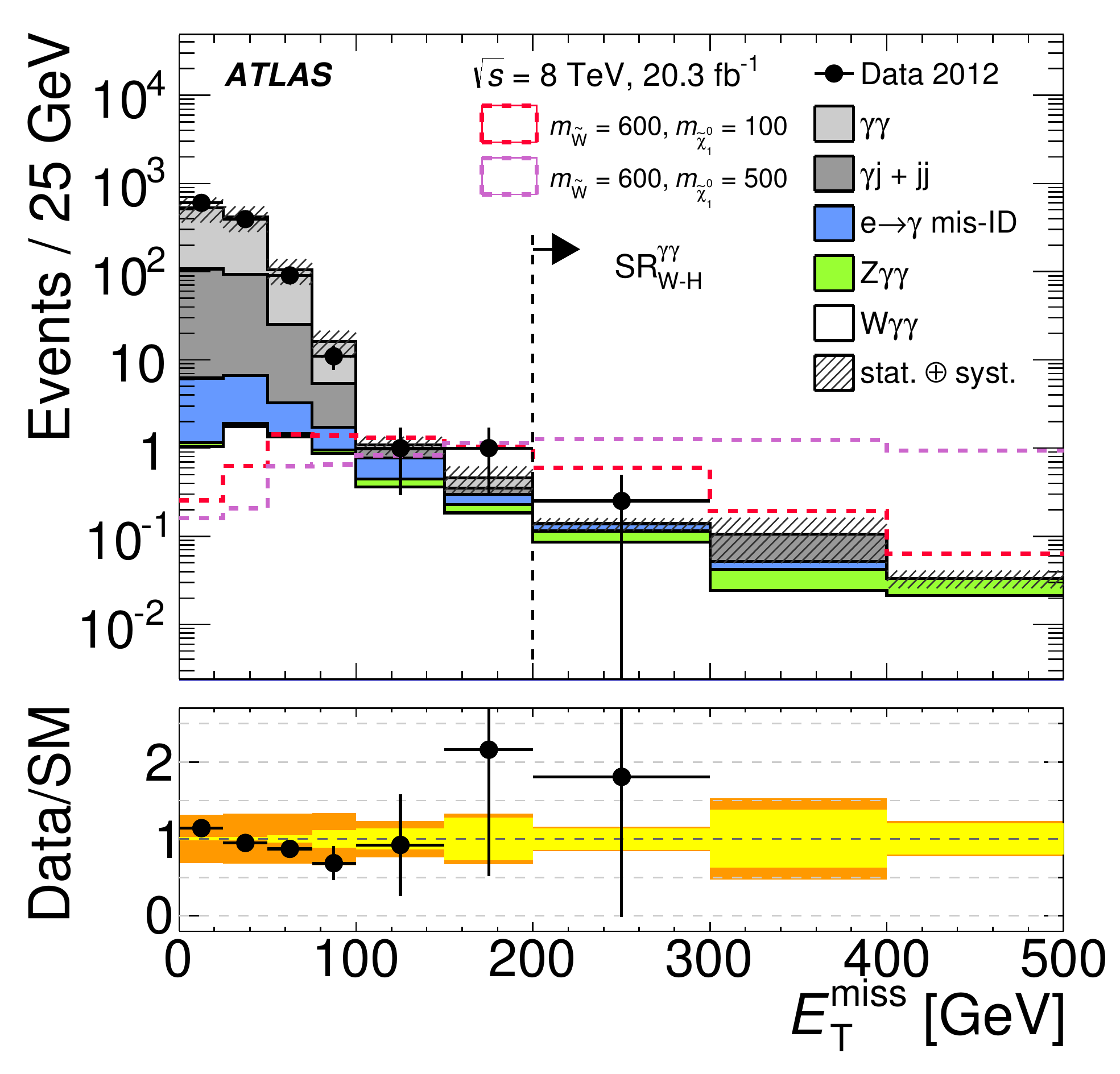} ~~
    \includegraphics[width=0.48\textwidth]{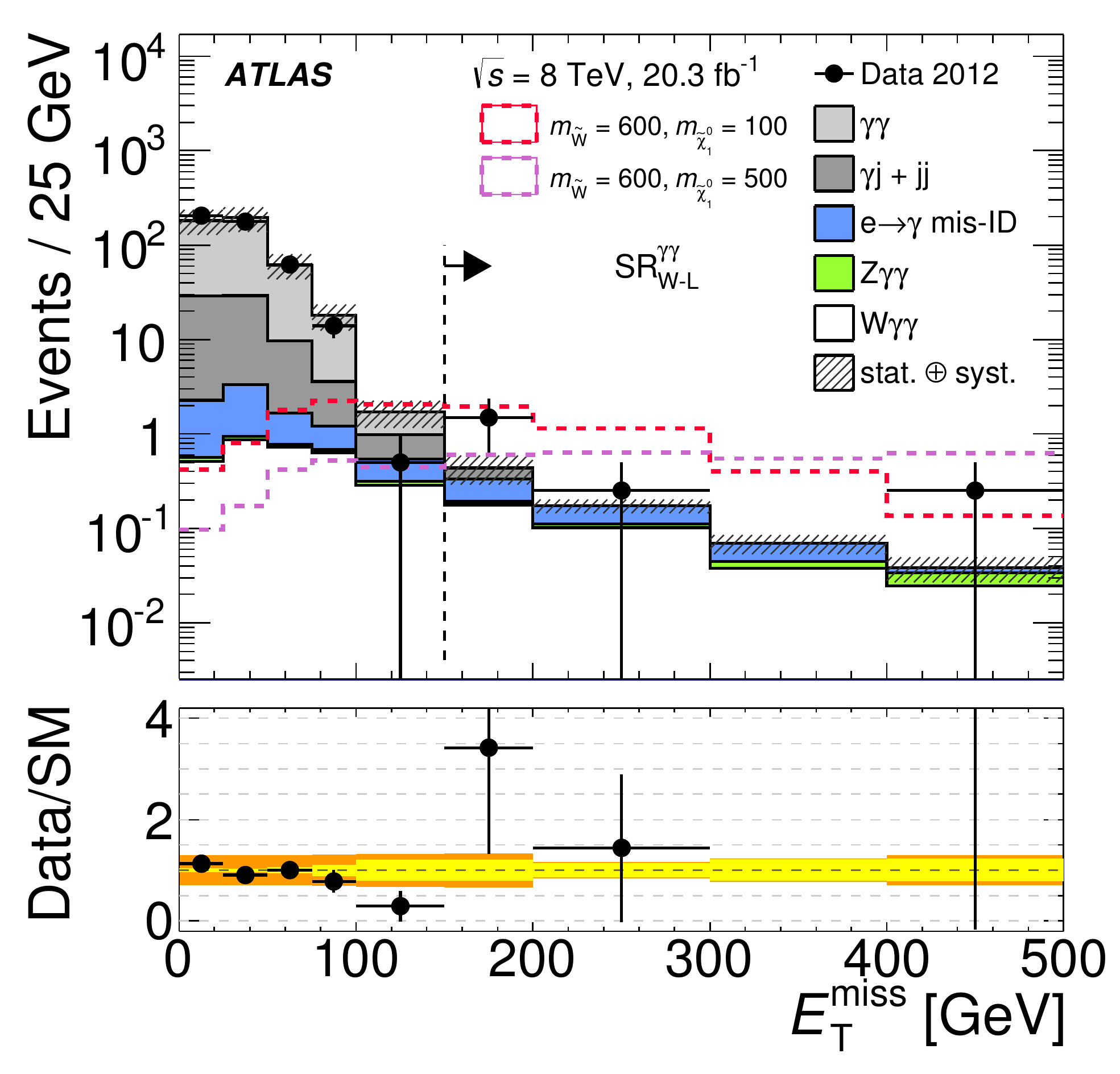}
  \end{center}
  \caption{Distribution of the missing transverse momentum \MET for the sample surviving all
requirements of the \BWH (left) and \BWL (right) selection except the \MET requirement itself. Overlain
are the expected SM backgrounds as a function of \MET, separated into
the various contributing sources. Also shown are the signal expectations
for the $(\mass{\wino},\mass{\neutralino}) = (600,100)$ GeV
and $(\mass{\wino},\mass{\neutralino}) = (600,500)$ GeV models.
The lower plots show the ratio of observed data to the combined SM expectation.
For these plots, the inner band represents the range of statistical uncertainty while
the outer band represents the combined statistical and systematic uncertainty.
Events outside the range of the displayed region are included in the highest-valued bin.
    \label{fig:diphoton_met}
  }
\end{figure}

\begin{figure}[tp]
  \begin{center}
    \includegraphics[width=0.48\textwidth]{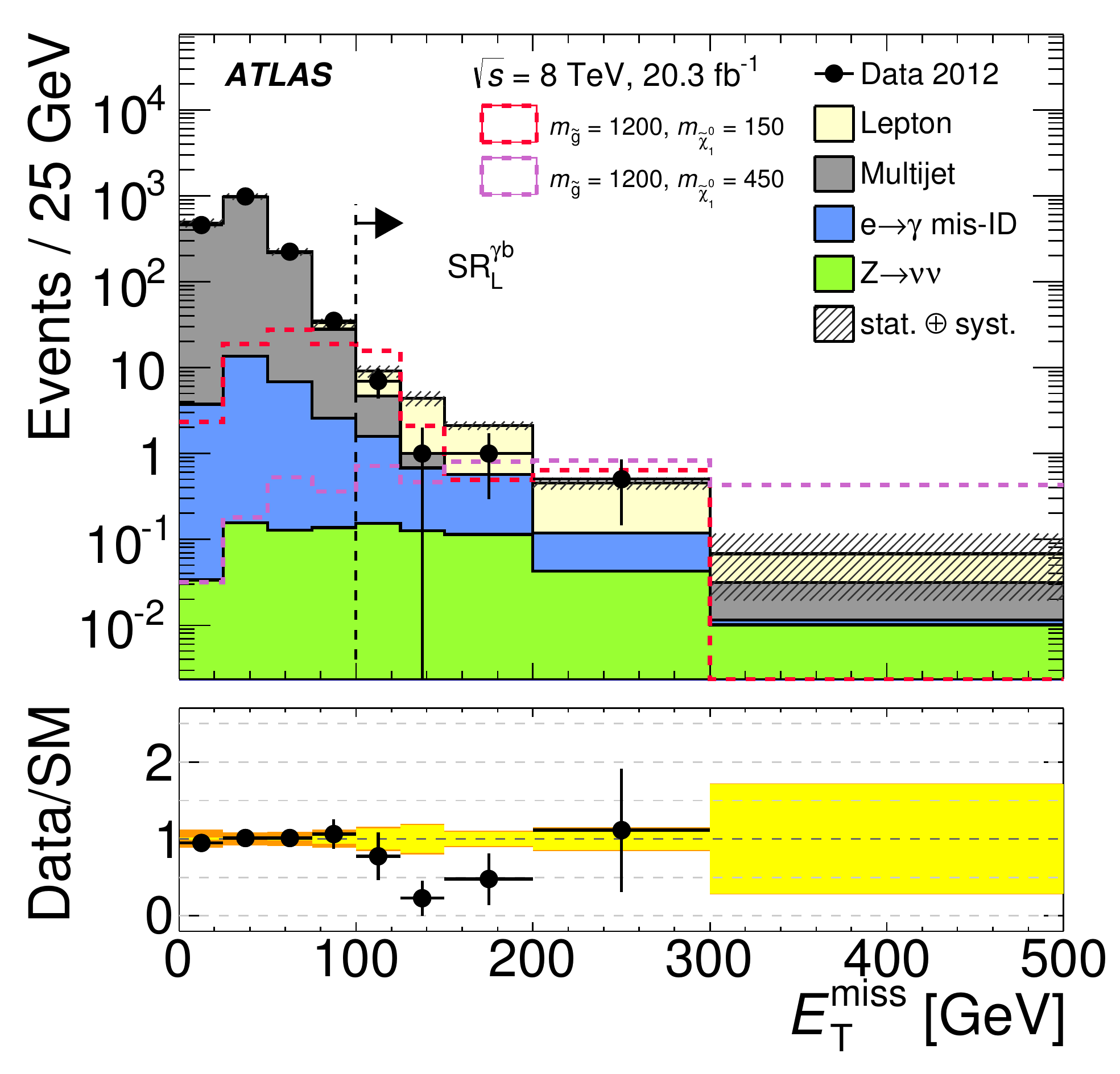} ~~
    \includegraphics[width=0.48\textwidth]{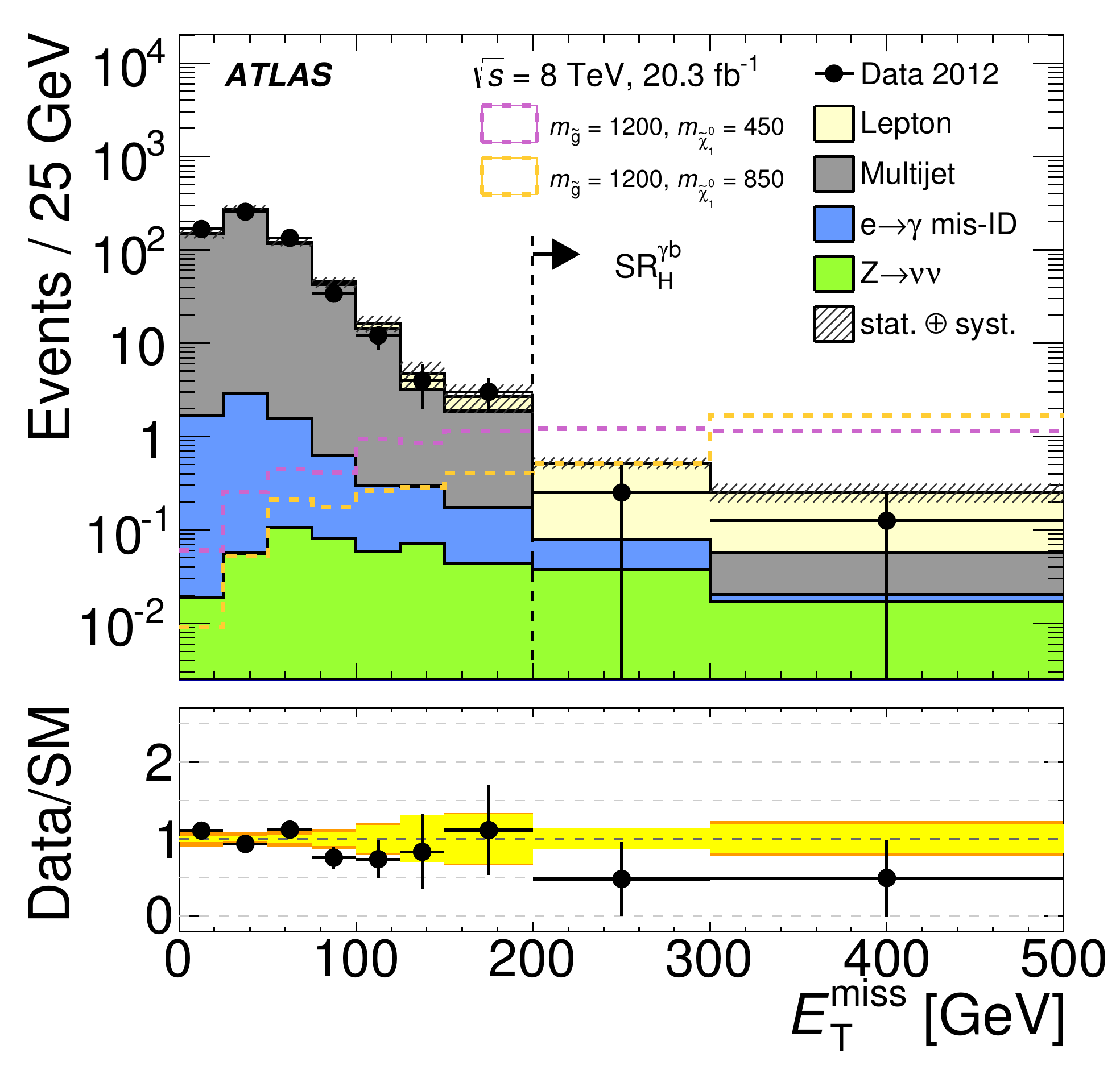}
  \end{center}
  \caption{\MET distribution for the sample surviving all
requirements of the \HBNL (left) and \HBNH (right) selection except the \MET requirement itself. Overlain
are the expected SM backgrounds as a function of \MET, separated into
the various contributing sources. Also shown are the signal expectations
for the $(\mass{\gluino},\mass{\neutralino})$ = (1200,150), (1200,450), and (1200,850)~GeV models.
The lower plots show the ratio of observed data to the combined SM expectation.
For these plots, the inner band represents the range of statistical uncertainty while
the outer band represents the combined statistical and systematic uncertainty.
Events outside the range of the displayed region are included in the highest-valued bin.
    \label{fig:photon_b_met}
  }
\end{figure}

\begin{figure}[tp]
  \begin{center}
  \includegraphics[width=0.48\textwidth]{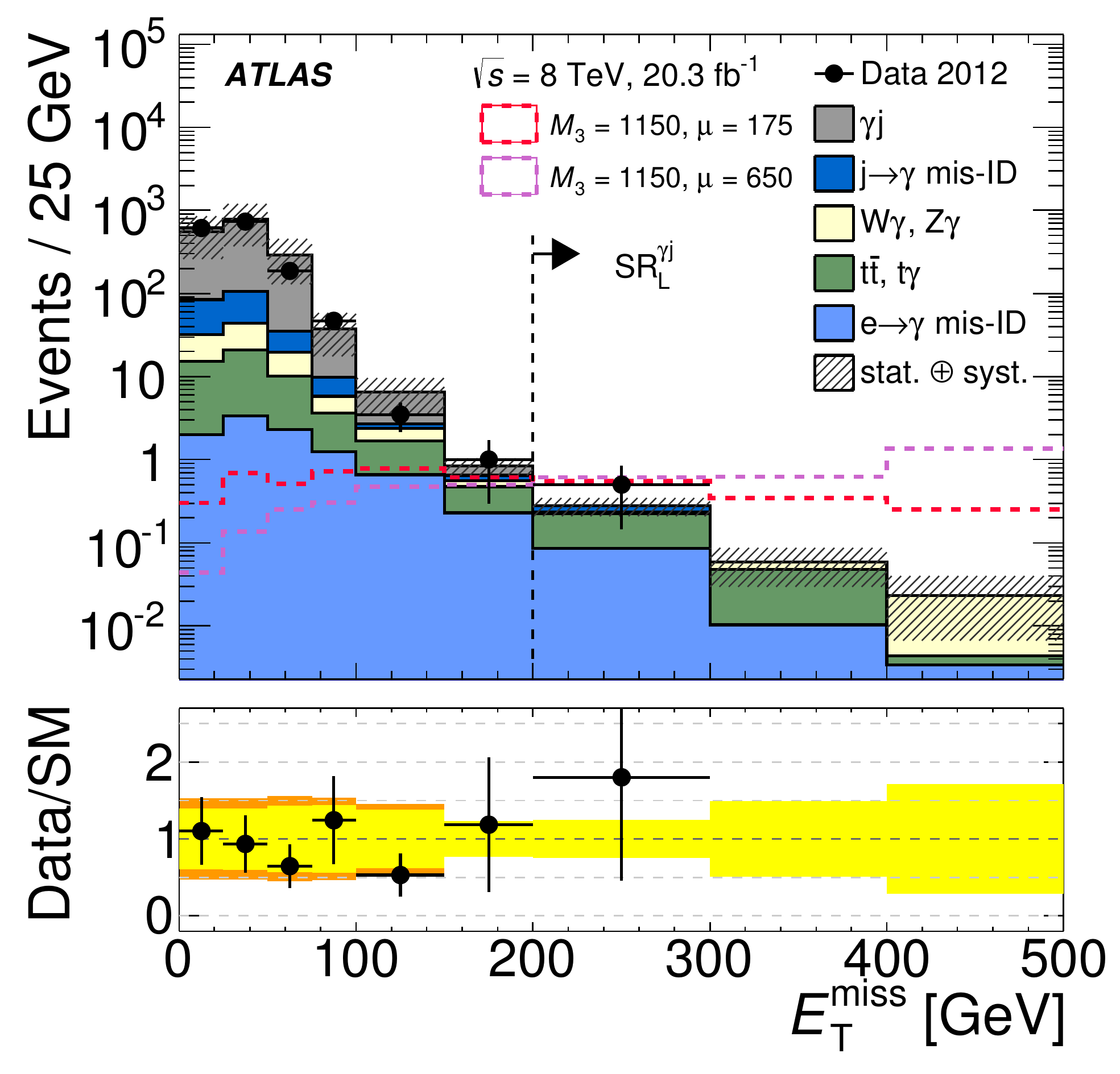} ~~
  \includegraphics[width=0.48\textwidth]{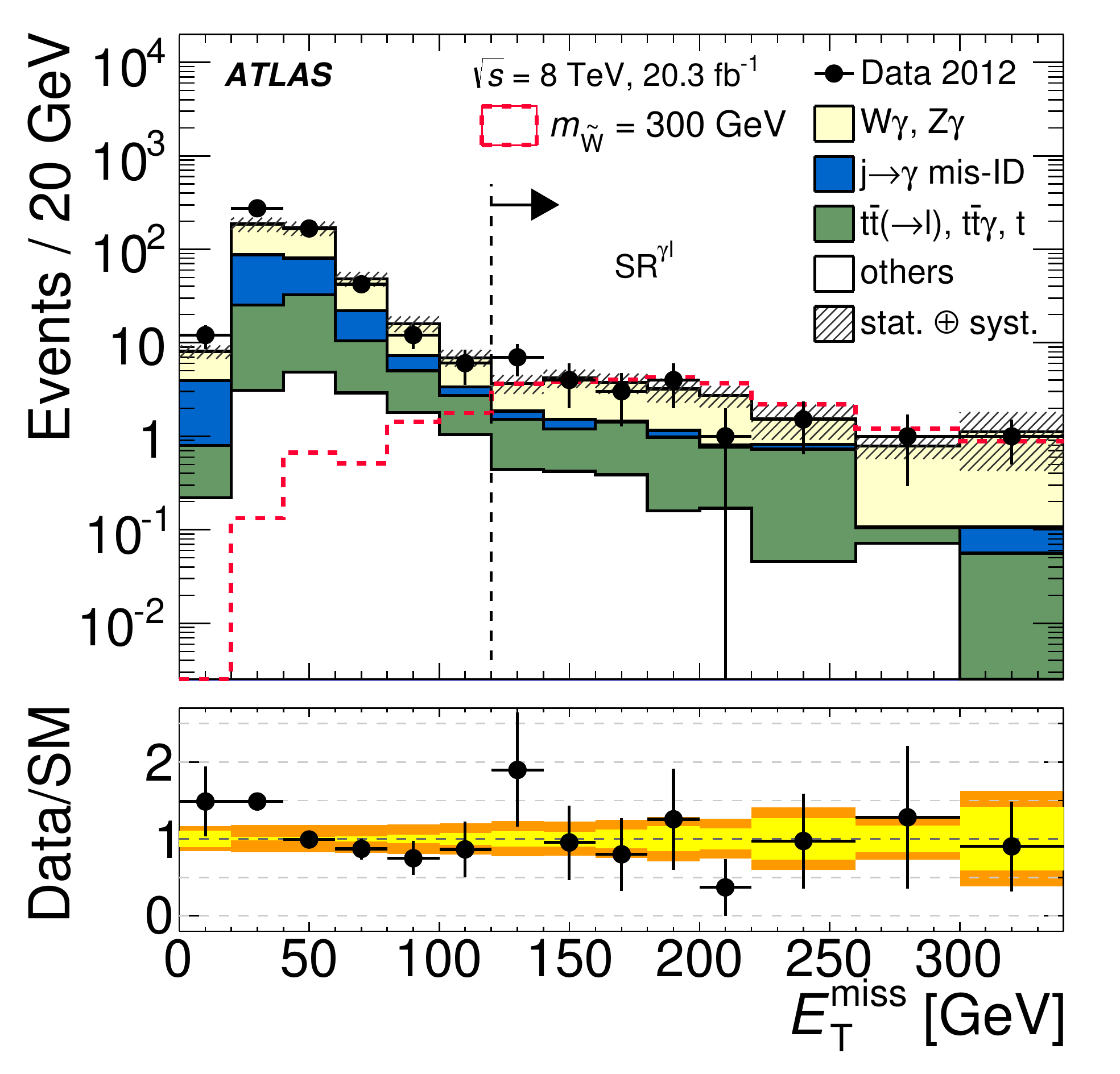}
  \end{center}
  \caption{
    (Left)~\MET distribution for the sample surviving all \HBPL requirements
    except the \MET requirement itself. Overlain
    are the expected SM backgrounds as a function of \MET, separated into
    the various contributing sources. Also shown are the signal expectations
    for the two points in the $M_3$--$\mu$ parameter space of the
    GGM model relevant to the photon+$j$ analysis.
    (Right)~\MET distribution for the combined 
    sample of events surviving all \WNLSPe and
    \WNLSPu requirements
    except the \MET requirement itself. Overlain
    are the expected SM backgrounds as a function of \MET, separated into
    the various contributing sources. Also shown is the signal expectation
    for the $\mass{\wino} = \unit[300]{GeV}$
    GGM model relevant to the photon+$\ell$ analysis.
    For both figures, the lower plot shows the ratio of observed data to the combined SM expectation,
    with the inner band representing the range of statistical uncertainty and
    the outer band the combined statistical
    and systematic uncertainty.
    Events outside the range of the displayed region are included in the highest-valued bin.
    }
    \label{fig:photon_l_met}
\end{figure}

Based on the numbers of observed events in the ten SRs and the
background expectation shown in Table~\ref{tab:mod_ind_lim},
95\% confidence-level (CL) upper limits are set for each SR on the number of events 
from any scenario of physics beyond the SM, using the
profile likelihood and $CL_s$ prescriptions~\cite{Read:2002hq}.
Uncertainties
on the background expectations are treated as
Gaussian-distributed nuisance parameters in the maximum-likelihood fit.
Assuming that no events due to physical processes beyond those
of the SM populate the various CRs used to estimate SR backgrounds,
observed 95\% CL limits on the number of such events vary between 3.0
(for the \BSL and \BSH SRs) and 14.2 (for the \WNLSPe SR).  
Taking into account the
integrated luminosity of \integLumiE
these number-of-event limits translate into 95\% CL upper
limits on the visible cross section for new physics, defined by the
product of cross section, branching fraction, acceptance and
efficiency for the different SR definitions. Correspondingly, 
the observed visible cross-section limits vary between 
0.15 and 0.70 fb.

By considering, in addition, the value and uncertainty of the
acceptance times efficiency of the selection requirements associated with
the various SRs,
as well as the NLO (+NLL) GGM
cross sections~\cite{Beenakker:1996ch,Kulesza:2008jb,Kulesza:2009kq,Beenakker:2009ha,Beenakker:2011fu},
which vary steeply with gluino and gaugino mass,
95\% CL lower limits may be set on the masses of these states in the
context of the various GGM scenarios explored in this study. 
For the diphoton, photon+$b$ and photon+$j$ analysis, the SR with the
best expected sensitivity at each simulated point in the parameter space
of the corresponding GGM model(s) is used to determine the
degree of exclusion of that model point. 
For the photon+$\ell$ analysis, the 95\% CL exclusion limits are derived from the combined likelihood
of the electron and muon channels, taking into account the correlation between the systematic
uncertainty estimates in the two channels.

For the diphoton analysis, \BSH is expected to provide the 
greatest sensitivity to the gluino-bino model for bino masses
above \unit[800]{GeV} and \BSL for bino masses below this.
For the wino-bino model, the similar transition point between
the use of \BWL and \BWH is found to be at \unit[350]{GeV}.
The resulting observed limits on the gluino and wino masses
are exhibited, as a function
of bino mass, for the diphoton analysis gluino and wino production models in
Figs.~\ref{fig:di_gluino_limits} and~\ref{fig:di_wino_limits}, respectively.
For the purpose of establishing these model-dependent limits,
for all four diphoton SRs both the normalization of the \Wgg-background estimate 
and the limit on the possible number of events from new physics
are extracted from a simultaneous fit to the SR and \Wgg control region,
although the signal contamination in the \Wgg control sample is appreciable
only for the low-bino-mass region of the wino-bino parameter space. 
Also shown for these two figures, as well as for the following two
figures (Figs.~\ref{fig:photon_b_limits} and~\ref{fig:gamma_j_limits}),
are the expected limits, including their 
statistical and background uncertainty ranges,
as well as observed limits for SUSY model cross sections $\pm 1$ standard deviation of theoretical
uncertainty from their central value. 
Conservatively choosing the $-1$ standard-deviation
observed contour, 95\% CL lower limits of
\GGMlimitG and \GGMlimitW 
are set by the diphoton analysis on the value of the gluino or wino mass, respectively,
for any value of the NLSP bino mass less than that
of the gluino (wino) mass.

Due to the discrete nature of the number-of-observed-events likelihood distribution in background-only
pseudoexperiments, when both the
expected number of observed events and its uncertainty are close to zero
the expected limit is dominated by the case of zero observed events.
This leads to a very narrow
one-standard-deviation range for the expected limit,
as observed for the expected-limit contour displayed in Fig.~\ref{fig:di_gluino_limits}.
In addition, because the observed number of
events is very close to the expected number of events for \BSH and \BSL, the expected
and observed limits are nearly identical in Fig.~\ref{fig:di_gluino_limits}.

\begin{figure}[tp]
  \begin{center}
    \includegraphics[width=0.80\textwidth]{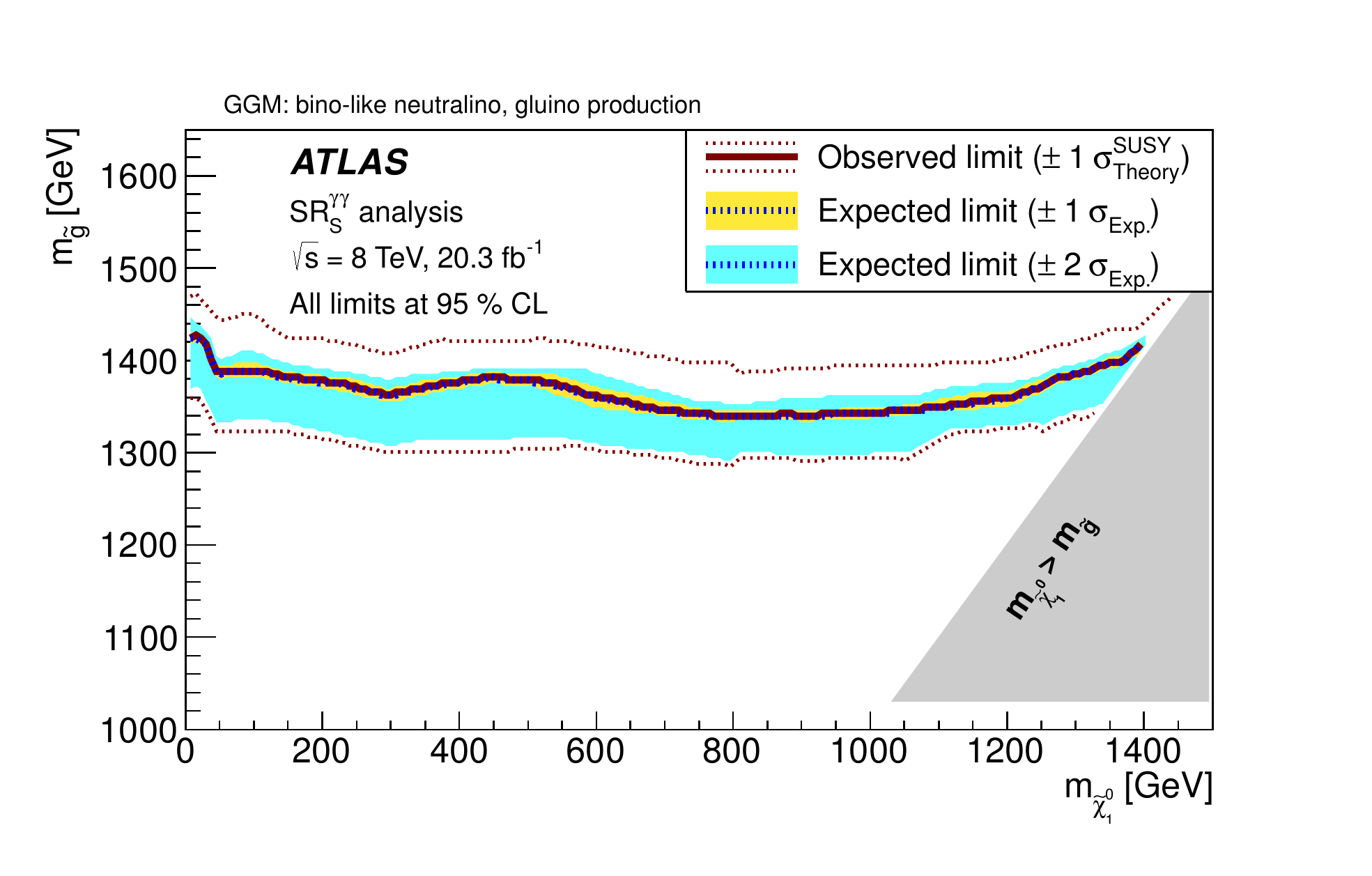}
    \caption{
Exclusion limits in the gluino-bino mass plane, using the \BSH analysis for
$\mass{\neutralino} \ge 800$ GeV and the \BSL analysis for
$\mass{\neutralino} < 800$ GeV.
Combinations of gluino and bino mass are excluded at 95\% CL 
in the area below the unbroken curve.
The observed limits are exhibited for the nominal SUSY model cross-section
expectation, as well as for a SUSY cross section increased and decreased
by 1 standard deviation of the cross-section systematic uncertainty.
Also shown is the expected limit, as well as the $\pm 1$ and $\pm 2$ standard-deviation
ranges of the expected limit. 
}
    \label{fig:di_gluino_limits}
  \end{center}
\end{figure}

\begin{figure}[tp]
  \begin{center}
    \includegraphics[width=0.80\textwidth]{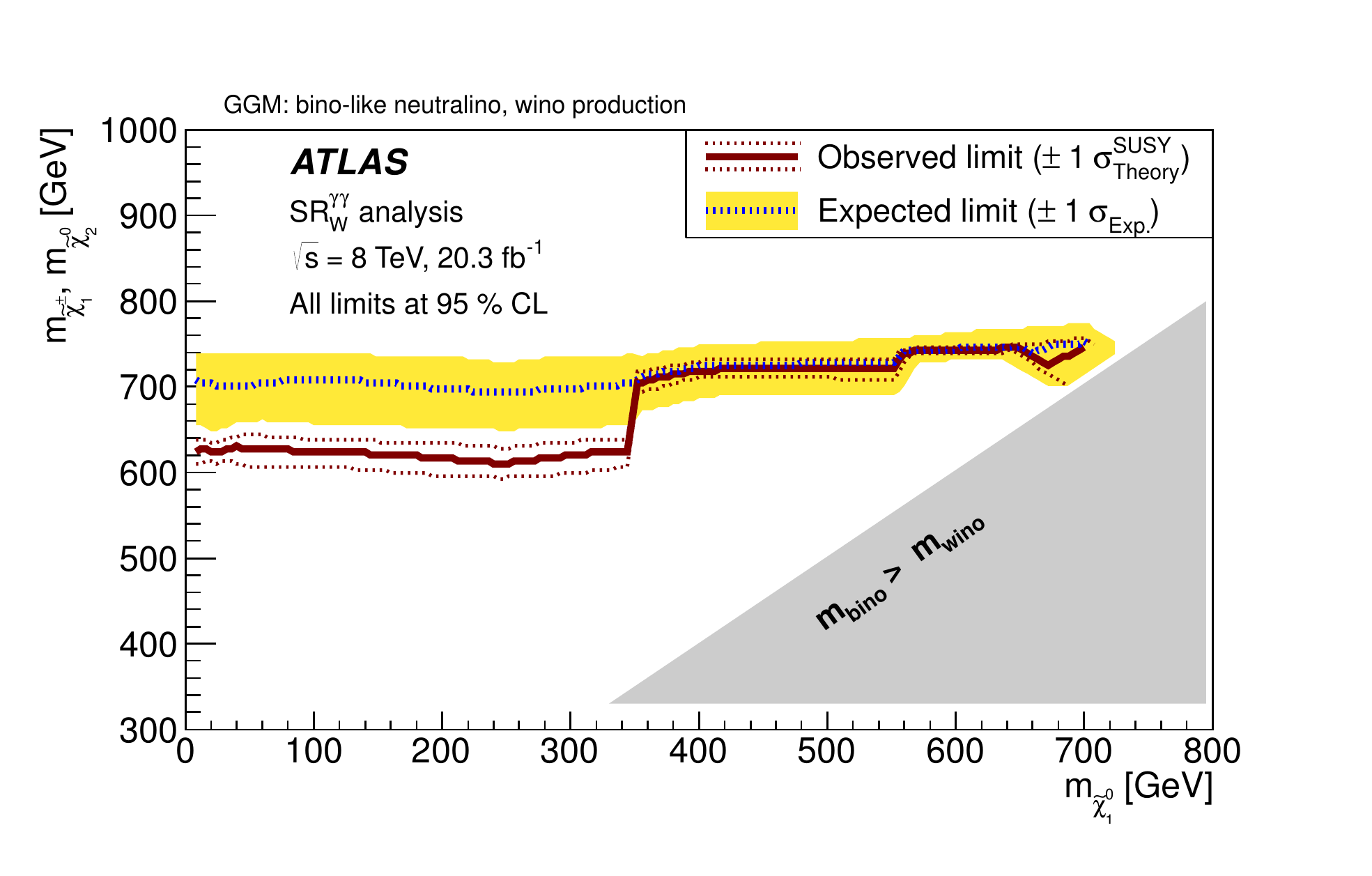}
    \caption{
Exclusion limits in the wino-bino
mass plane, using the \BWH analysis for
$\mass{\neutralino} \ge 350$ GeV and the \BWL analysis for
$\mass{\neutralino} < 350$ GeV.
The vertical axis represents wino mass while the horizontal axis represents bino mass.
The observed limits are exhibited for the nominal SUSY model cross-section
expectation, as well as for a SUSY cross section increased and decreased
by 1 standard deviation of the cross-section systematic uncertainty.
Also shown is the expected limit, along with its $\pm 1$
standard-deviation range.
The discontinuity at $\mass{\neutralino} = 350$ GeV is due to the
switch between the use of the \BWL and \BWH analyses, the former
of which exhibits a small excess of observed events relative to
the expected SM background.
}
    \label{fig:di_wino_limits}
  \end{center}
\end{figure}

\begin{figure}[tp]
  \begin{center}
    \includegraphics[width=0.80\textwidth]{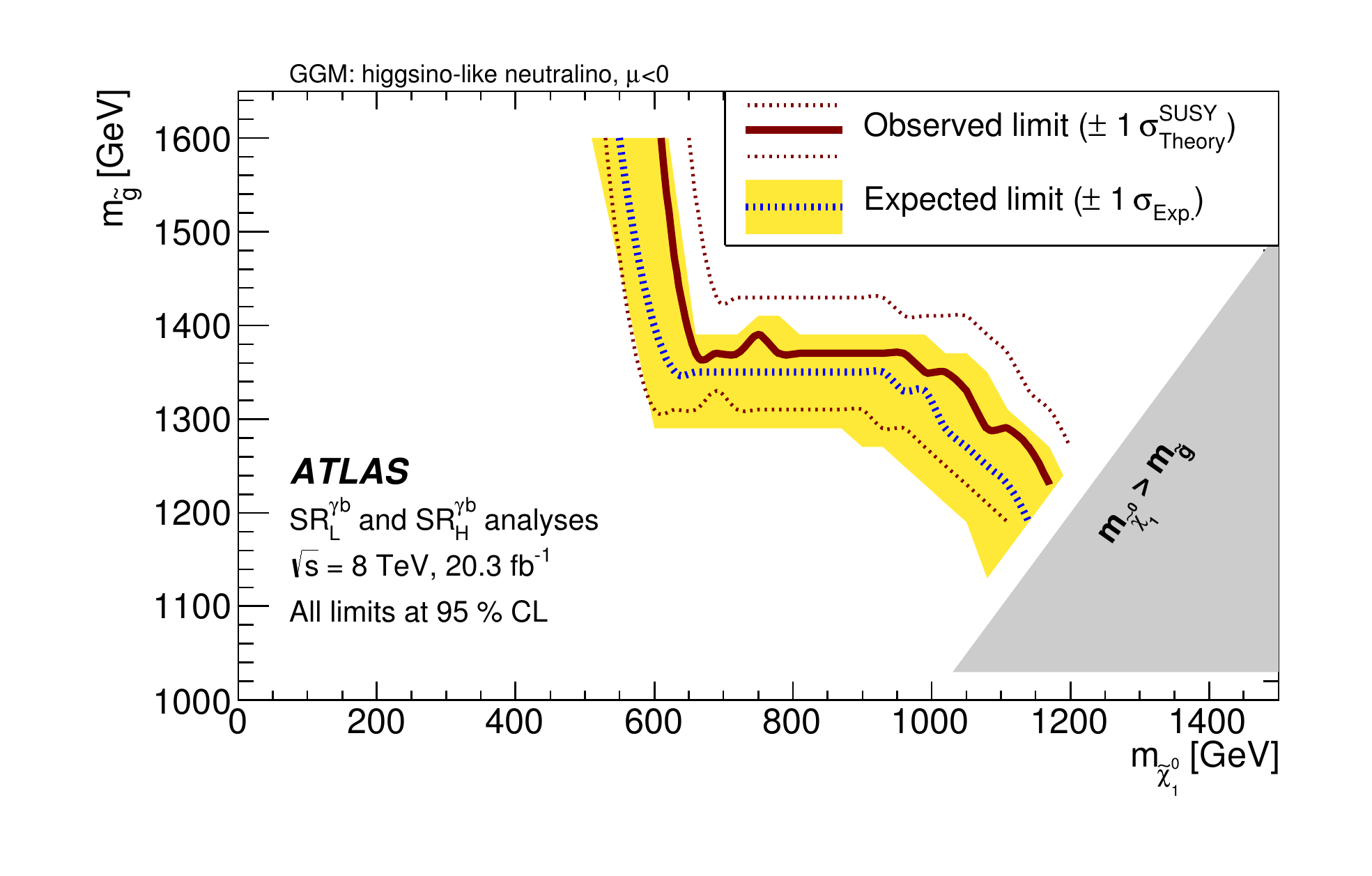}
    \caption{
Exclusion limits in the gluino-neutralino mass plane, for the higgsino-bino
GGM model with $\mu < 0$, using the merged (see text) \HBNL and \HBNH analyses.
The observed limits are shown for the nominal SUSY model cross-section
expectation, as well as for a SUSY cross section increased and decreased
by 1 standard deviation of the cross-section systematic uncertainty.
The expected limit is also shown, along with its $\pm 1\sigma$ range.
For NLSP masses below approximately \unit[450]{GeV}, the
onset of the direct production of gaugino states
makes the analysis insensitive to the value of the gluino mass.}
    \label{fig:photon_b_limits}
  \end{center}
\end{figure}

For the photon+$b$ analysis, limits are set in the two-dimensional plane of gluino 
and $\neutralino$ mass for the higgsino-bino GGM model with a negative value of
the $\mu$ parameter. For NLSP masses near the 95\% CL exclusion contour, \HBNL is 
expected to provide greater sensitivity for 
NLSP masses below approximately \unit[600]{GeV}, and so is made use of in this region;
above that, \HBNH is used to establish the degree of exclusion of 
points in the GGM model space.
The resulting observed exclusion contour is shown in Fig.~\ref{fig:photon_b_limits}. 
Again choosing the $-1$ standard-deviation
observed contour, in the context of this GGM model a conservative lower 
limit of \unit[1300]{GeV} is established
for the gluino mass over much of the range of the higgsino-bino NLSP
mass. For NLSP masses above \unit[1000]{GeV} the sensitivity lessens
due to the restriction of the phase space for producing an
energetic $b$-jet, while for NLSP masses below \unit[600]{GeV}, the
onset of the direct production of gaugino states begins to 
make the analysis insensitive to the value of the gluino mass.

For the photon+$j$ analysis, limits are set in the two-dimensional plane of the GGM
parameters $\mu$ and $M_3$ for the higgsino-bino GGM model with a positive value of
the $\mu$ parameter. For values of $\mu$ near the 95\% CL exclusion contour, \HBPL is
expected to provide a greater sensitivity for
NLSP masses below approximately \unit[900]{GeV}, and so is made use of in this region;
above that, \HBPH is used to establish the degree of exclusion of GGM model-space points.
The resulting observed exclusion contour is shown in Fig.~\ref{fig:gamma_j_limits}. 
Again choosing the $-1$ standard-deviation
observed contour, in the context of this GGM model a conservative lower
limit of \unit[1140]{GeV} is established
for the gluino mass parameter $M_3$ over much of the range of the $\mu$ parameter.
For values of $M_3$ close to the value of $\mu$ for which the gluino mass
approaches that of the higgsino-bino NLSP, the sensitivity of the analysis
lessens due to the restriction of phase space for producing multiple high-$\pt$ jets.

For the photon+$\ell$ analysis, a limit is set on the wino mass, the
single free parameter of the wino-NLSP
model. Fig.~\ref{fig:ppl_limits} shows the observed limit on the cross
section for wino production in this model, as well as the
corresponding expected limit with $\pm 1$ and $\pm 2$
standard-deviation uncertainty bands. Also shown is the cross section
as a function of wino mass, with its $\pm 1$ standard-deviation range.
In the context of this wino-NLSP model, conservatively choosing the
$-1$ standard-deviation cross-section contour leads to an exclusion of
GGM winos in the the range $124 < M_{\wino} < \unit[361]{GeV}$; for
$M_{\wino} < \unit[124]{GeV}$ the signal contamination in the
$W\gamma$ CR becomes too large to permit a reliable estimate of the
$W\gamma$ background.  These limits are based on the direct production
of the wino NLSP in the limit where squark masses are infinite, and
are independent of gluino mass.

\begin{figure}[tp]
  \begin{center}
    \includegraphics[width=0.80\textwidth]{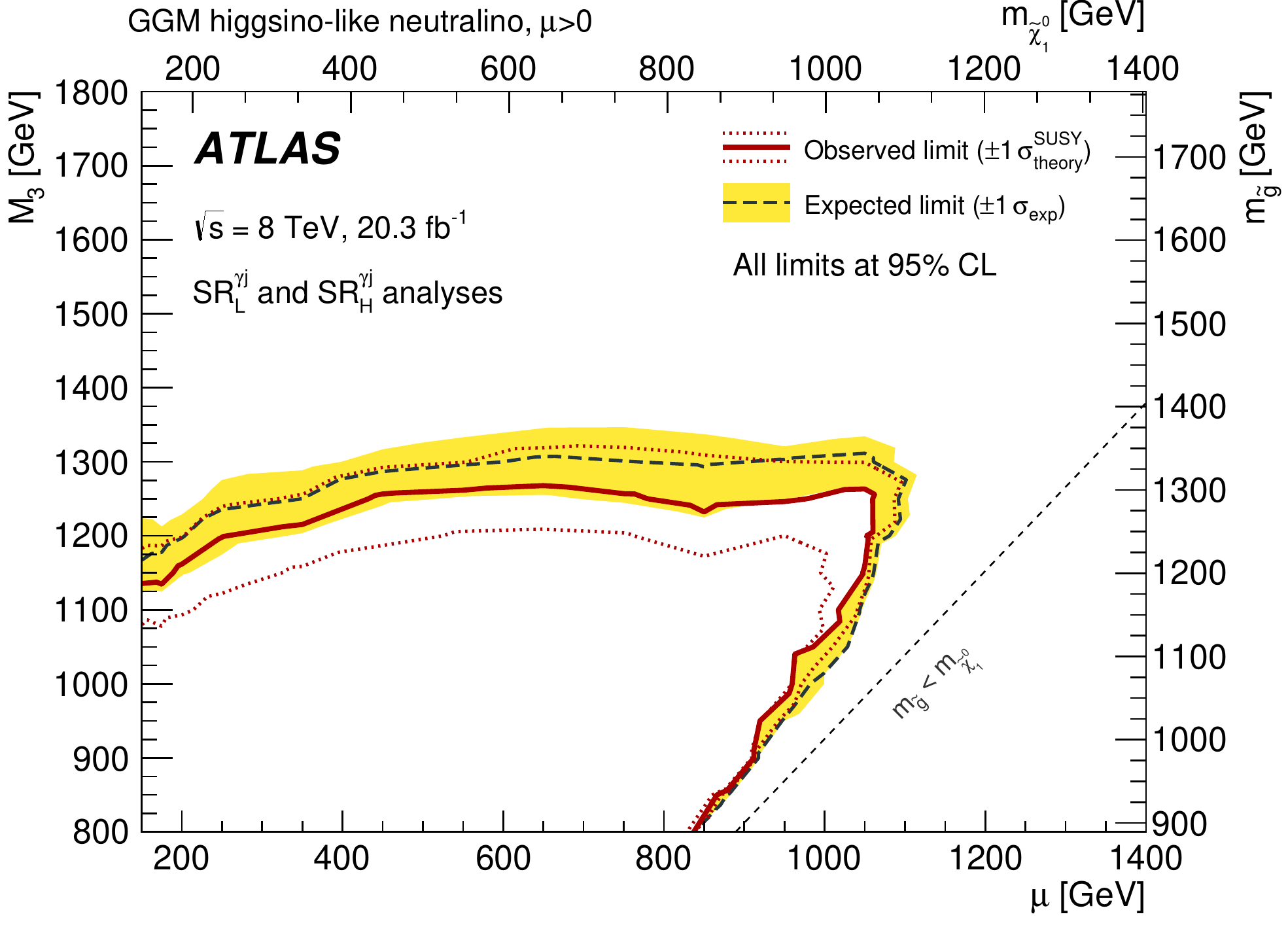}
    \caption{
Exclusion limits imposed by the photon+$j$ analysis
in the two-dimensional plane of the GGM parameters $M_3$ and $\mu$,
for the higgsino-bino GGM model with $\mu > 0$,
using the merged (see text) \HBPL and \HBPH analyses.
The observed limits are shown for the nominal SUSY model cross-section
expectation, as well as for a SUSY cross section increased and decreased
by 1 standard deviation of the cross-section systematic uncertainty.
The expected limit is also shown, along with its $\pm 1$ standard-deviation range.
Values of $M_3$ below \unit[1100]{GeV} are excluded for most values
of the $\mu$ parameter, although a significant region corresponding
to the case for which the gluino mass is close to that
of the lightest neutralino masses remains unexcluded due to the
requirements of one or more jets arising from the gluino decay.
The top and right axes represent the corresponding values of the lightest
neutralino and gluino masses, respectively.
}
    \label{fig:gamma_j_limits}
  \end{center}
\end{figure}

\begin{figure}[tp]
  \begin{center}
    \includegraphics[width=0.80\textwidth]{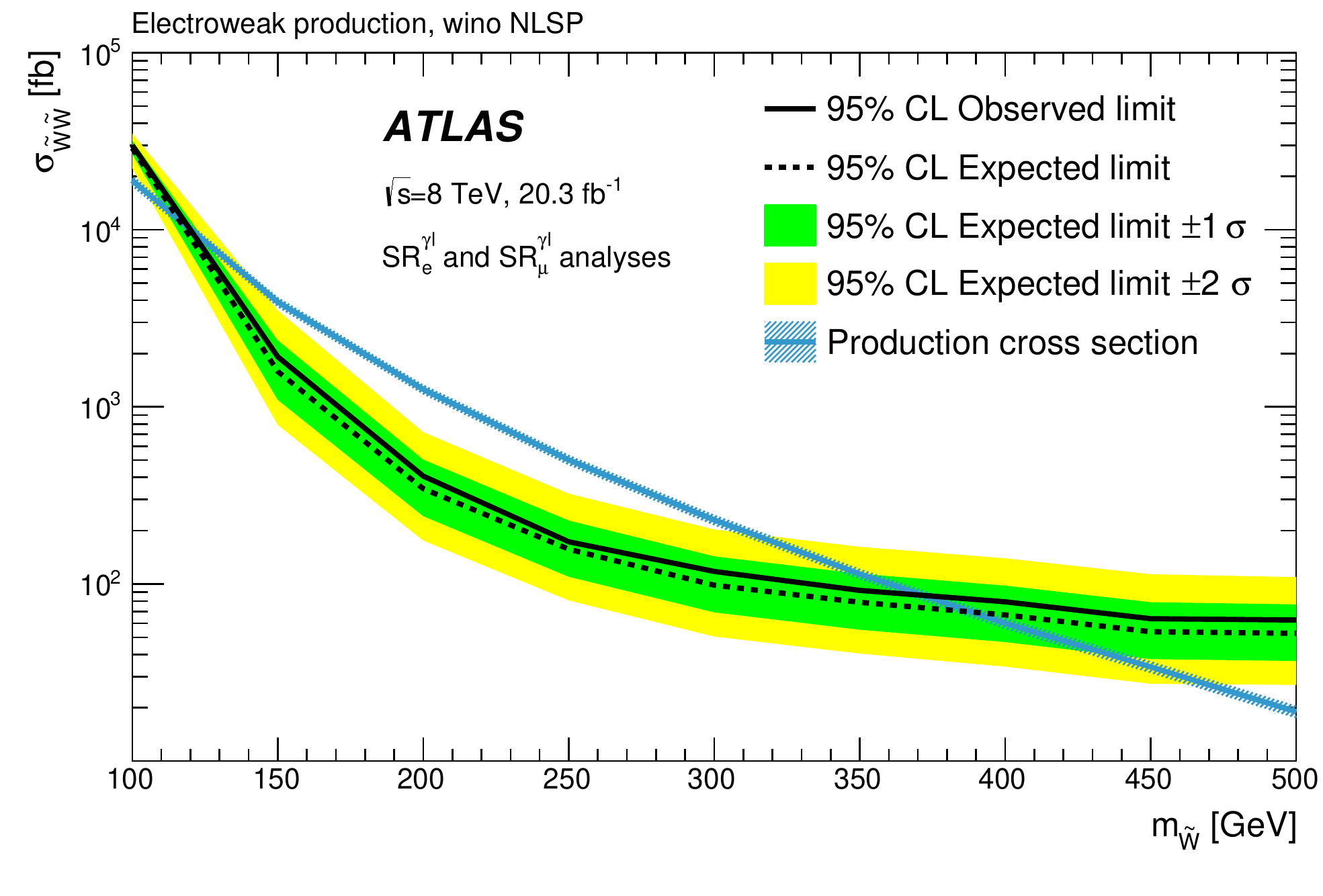}
    \caption{Contour of exclusion in wino production 
cross section
from the photon+$\ell$ analysis, as
a function of the wino mass parameter $M_{\tilde{W}}$.
The expected limit is shown along with its $\pm 1$ and $\pm 2$ standard-deviation ranges.
Also shown is the total cross section for the production of
\wino pairs, again as a function of $M_{\tilde{W}}$.}
    \label{fig:ppl_limits}
  \end{center}
\end{figure}

\FloatBarrier

\section{Conclusion}
\label{sec:conclusion}

Making use of \integLumi of $pp$ collision data at $\sqrt{s} = 8$ TeV recorded by the ATLAS
detector at the LHC, a search is performed for photonic
signatures of new physics associated with significant \met.
Four experimental signatures are explored, each involving at least one energetic isolated 
final-state photon in association with significant \met, and used to
search for evidence for several GGM SUSY scenarios. 
No significant excess of events over the SM expectation
is observed in any of the searches and so limits are
set on possible contributions of new physics. Model-independent limits are set
on the numbers of events from new physics and the associated visible
cross section. Model-dependent limits are set on the masses of SUSY particles
or on mass parameters associated with the various GGM scenario models.
 
A diphoton signature is used to explore both strongly and weakly produced SUSY states with
a decay chain proceeding through a binolike NLSP. In the context of these
models, lower limits of
\GGMlimitG and \GGMlimitW are set on the masses of a degenerate octet of gluinos and
a degenerate set of winos, respectively, for any value of the bino mass less
than the mass of these produced states. A photon-plus-$b$-jet signature is
used to search for a scenario in which the GGM NLSP is a higgsino-bino
admixture with a roughly equal branching fraction to photons and to the
SM Higgs boson. In the context of this model, a lower limit of \unit[1260]{GeV} is established
for the gluino mass over much of the range of the higgsino-bino NLSP
mass; for NLSP masses below approximately \unit[450]{GeV}, the
onset of the direct production of gaugino states 
makes the analysis insensitive to the value of the gluino mass.
A photon-plus-jet signature is
used to search for an alternative scenario for which the GGM NLSP is a higgsino-bino
admixture with a roughly equal branching fraction to photons and to the
SM $Z$ boson. In the context of this model, a lower
limit of \unit[1140]{GeV} is established
for the gluino mass parameter $M_3$ over much of the range of the higgsino mass parameter $\mu$.
Finally, a photon-plus-lepton signature is used to search
for a scenario for which the GGM NLSP is a degenerate set of three wino states.
Based on the possible direct production of these states, in the limit of infinite
squark mass GGM winos are excluded in the 
range $124 < M_{\wino} < \unit[361]{GeV}$, independent of the gluino mass.

\section*{Acknowledgments}

We thank CERN for the very successful operation of the LHC, as well as the
support staff from our institutions without whom ATLAS could not be
operated efficiently.

We acknowledge the support of ANPCyT, Argentina; YerPhI, Armenia; ARC,
Australia; BMWFW and FWF, Austria; ANAS, Azerbaijan; SSTC, Belarus; CNPq and FAPESP,
Brazil; NSERC, NRC and CFI, Canada; CERN; CONICYT, Chile; CAS, MOST and NSFC,
China; COLCIENCIAS, Colombia; MSMT CR, MPO CR and VSC CR, Czech Republic;
DNRF, DNSRC and Lundbeck Foundation, Denmark; EPLANET, ERC and NSRF, European Union;
IN2P3-CNRS, CEA-DSM/IRFU, France; GNSF, Georgia; BMBF, DFG, HGF, MPG and AvH
Foundation, Germany; GSRT and NSRF, Greece; RGC, Hong Kong SAR, China; ISF, MINERVA, GIF, I-CORE and Benoziyo Center, Israel; INFN, Italy; MEXT and JSPS, Japan; CNRST, Morocco; FOM and NWO, Netherlands; BRF and RCN, Norway; MNiSW and NCN, Poland; GRICES and FCT, Portugal; MNE/IFA, Romania; MES of Russia and NRC KI, Russian Federation; JINR; MSTD,
Serbia; MSSR, Slovakia; ARRS and MIZ\v{S}, Slovenia; DST/NRF, South Africa;
MINECO, Spain; SRC and Wallenberg Foundation, Sweden; SER, SNSF and Cantons of
Bern and Geneva, Switzerland; NSC, Taiwan; TAEK, Turkey; STFC, the Royal
Society and Leverhulme Trust, United Kingdom; DOE and NSF, United States of
America.

The crucial computing support from all WLCG partners is acknowledged
gratefully, in particular from CERN and the ATLAS Tier-1 facilities at
TRIUMF (Canada), NDGF (Denmark, Norway, Sweden), CC-IN2P3 (France),
KIT/GridKA (Germany), INFN-CNAF (Italy), NL-T1 (Netherlands), PIC (Spain),
ASGC (Taiwan), RAL (UK) and BNL (USA) and in the Tier-2 facilities
worldwide.

\providecommand{\href}[2]{#2}\begingroup\raggedright\endgroup

\newpage
\begin{flushleft}
{\Large The ATLAS Collaboration}

\bigskip

G.~Aad$^{\rm 85}$,
B.~Abbott$^{\rm 113}$,
J.~Abdallah$^{\rm 151}$,
O.~Abdinov$^{\rm 11}$,
R.~Aben$^{\rm 107}$,
M.~Abolins$^{\rm 90}$,
O.S.~AbouZeid$^{\rm 158}$,
H.~Abramowicz$^{\rm 153}$,
H.~Abreu$^{\rm 152}$,
R.~Abreu$^{\rm 116}$
% ...
\bigskip
\\
% ...
$^{11}$ Institute of Physics, Azerbaijan Academy of Sciences, Baku, Azerbaijan\\
% ...
$^{85}$ CPPM, Aix-Marseille Universit{\'e} and CNRS/IN2P3, Marseille, France\\
% ...
$^{90}$ Department of Physics and Astronomy, Michigan State University, East Lansing MI, United States of America\\
% ...
$^{107}$ Nikhef National Institute for Subatomic Physics and University of Amsterdam, Amsterdam, Netherlands\\
% ...
$^{113}$ Homer L. Dodge Department of Physics and Astronomy, University of Oklahoma, Norman OK, United States of America\\
% ...
$^{116}$ Center for High Energy Physics, University of Oregon, Eugene OR, United States of America\\
% ...
$^{151}$ Institute of Physics, Academia Sinica, Taipei, Taiwan\\
$^{152}$ Department of Physics, Technion: Israel Institute of Technology, Haifa, Israel\\
$^{153}$ Raymond and Beverly Sackler School of Physics and Astronomy, Tel Aviv University, Tel Aviv, Israel\\
% ...
$^{158}$ Department of Physics, University of Toronto, Toronto ON, Canada\\
% ...
\end{flushleft}


\end{document}